\documentclass[sigplan,10pt]{acmart}
\usepackage[ruled,vlined]{algorithm2e}
\usepackage{amsmath}
\usepackage{makecell}
\usepackage{enumitem}
\usepackage{multicol}
\usepackage{multirow}
\usepackage{threeparttable}
\usepackage{hyperref}
\usepackage{url}
\usepackage{xcolor}     
\usepackage{colortbl}   
\usepackage{algorithm2e}
\newcommand{\hlstage}[2]{%
  \colorbox{#1}{%
    \parbox{\dimexpr 0.8\linewidth-\fboxsep\relax}{#2}%
  }%
}

\usepackage{tikz}
\newcommand{\stepnum}[1]{%
  \tikz[baseline=(n.base)]{
    \node[shape=circle, fill=black, text=white,
          inner sep=0pt, minimum size=0.95em,
          font=\sffamily\bfseries\footnotesize] (n) {#1};
  }%
}

\AtBeginDocument{%
  }

\copyrightyear{2027}
\acmYear{2027}
\setcopyright{cc}
\setcctype{by}
\acmConference[EuroSys '27]{22nd European Conference on Computer Systems}{April 19--23, 2027}{Rabat, Morocco}
\acmBooktitle{22nd European Conference on Computer Systems (EuroSys '27), April 19--23, 2027, Rabat, Morocco}
\acmDOI{10.1145/3842654.3848583}
\acmISBN{979-8-4007-2971-3/2027/04}

\acmBadgeR[https://www.acm.org/publications/policies/artifact-review-and-badging-current]{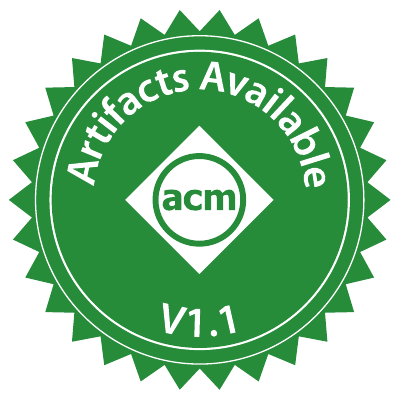}
\acmBadgeR[https://www.acm.org/publications/policies/artifact-review-and-badging-current]{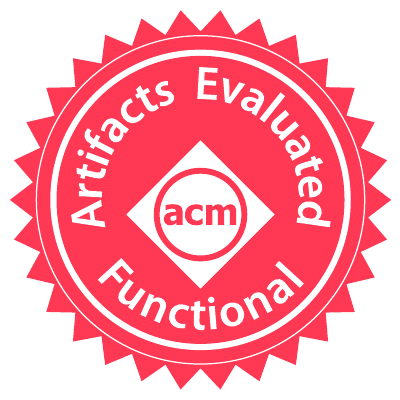}
\begin{document}

\title[\textsc{RapidMoE}: Exploiting Cross-Asymmetry via Adaptive Residual Offloading for Large-Scale MoE Inference]{\textsc{RapidMoE}: Exploiting Cross-Asymmetry via Adaptive Residual Offloading for Large-Scale MoE Inference}

\author{Wenxun Wang}
\orcid{0009-0007-1999-6441}
\affiliation{%
  \institution{Tsinghua University, BNRist}
  \city{Beijing}
  \country{China}
}
\email{wx-wang23@mails.tsinghua.edu.cn}

\author{Likai Ma}
\orcid{0009-0000-9297-0503}
\affiliation{%
  \institution{Tsinghua University}
  \city{Beijing}
  \country{China}}
\email{mlk22@mails.tsinghua.edu.cn}

\author{Zongle Huang}
\orcid{0009-0002-4557-3163}
\affiliation{%
  \institution{Tsinghua University, BNRist}
  \city{Beijing}
  \country{China}
}
\email{huangzl23@mails.tsinghua.edu.cn}

\author{Chen Tang}
\orcid{0000-0002-3411-0078}
\affiliation{%
 \institution{Tsinghua University}
\city{Beijing}
\country{China}
  }
\email{tangc20@mails.tsinghua.edu.cn}

\author{Yongpan Liu}
\orcid{0000-0002-4892-2309}
\authornote{Corresponding Author.}
\affiliation{%
  \institution{Tsinghua University, BNRist}
\city{Beijing}
\country{China}
  }
\email{ypliu@tsinghua.edu.cn}

\renewcommand{\shortauthors}{Wenxun Wang et al.}

\begin{abstract}
The widespread adoption of Mixture-of-Experts (MoE) has created a growing need for deployment on heterogeneous platforms. However, it exposes a fundamental mismatch between the algorithmic demands of large-scale MoE and the disparate characteristics of hardware.
Existing CPU-GPU hybrid inference systems fail to resolve this as they either encounter PCIe bandwidth bottlenecks when loading experts to GPUs, or rely heavily on CPU computation.
Consequently, this leads to low resource utilization and inevitable violations of fixed latency budgets as parameters scale.

In this paper, we identify and exploit Cross-Asymmetry—a structural alignment between the algorithmic workload skew of MoE routing and the physical disparity of heterogeneous hardware. 
To this end, we introduce \textsc{RapidMoE}, a residual offloading system for efficient large-scale MoE inference.
We propose how \textsc{RapidMoE} leverages a residual‑split framework to enable offloading paradigm shift from expert-level to bit-level, which unfolds across three key dimensions: (1) data representation, enabling compact and decoupled storage; (2) routing strategy, partitioning computation into dual paths aligned with hardware capabilities; (3) execution parallelism, scheduling a balanced storage-compute workload across devices.
We further employ a novel Unified Multi-Level Importance Arbitration to adaptively adjust the critical expert set at runtime, ensuring the accuracy-latency Pareto frontier.
These innovations exploit inherent cross-asymmetry, fundamentally breaking the algorithm-hardware misalignment. Experimental results show that \textsc{RapidMoE} achieves up to 3.5$\times$ speedup in decoding and 2.1$\times$ speedup in prefill compared to state-of-the-art (SOTA) offloading systems. 

\end{abstract}



\maketitle

\section{Introduction}
\begin{figure}[t]
\centering
\includegraphics[width=0.49\textwidth,height=0.20\textwidth]{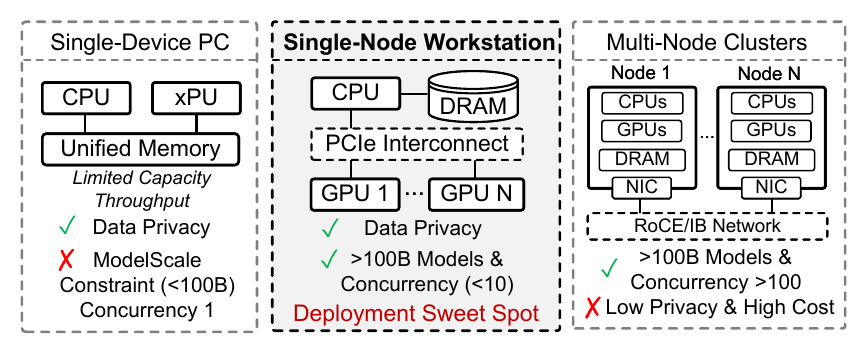}
  \caption{Comparison of deployment scenarios between single-device PC, multi-node clusters and single-node workstation.}
  \label{fig:workstation}
\end{figure}

The advancement of MoEs has spurred their adoption \cite{llm1,llm-survey,gpt4,llama}, achieving advanced capabilities with large-scale models such as DeepSeek-V3 \cite{deepseek-v3} and Qwen3-235B \cite{qwen3}. This evolution is driving a shift toward the deployment of large-scale MoEs on single-node heterogeneous platforms \cite{ktransformers}. 
As shown in \autoref{fig:workstation}, cloud clusters provide capacity but compromise privacy and cost-efficiency, whereas edge devices guarantee privacy but lack the memory and throughput for serving large-scale models \cite{llm-privacy}. Bridging this gap, the workstation—typically equipped with abundant host DRAM and limited GPU resources (e.g., RTX 4090 or A100)—has become a cost-effective deployment alternative. It satisfies privacy and capacity demands while offering potential to maintain strict Time Per Output Token (TPOT) for interactive streaming under low-concurrency (< 10).

\begin{figure}[t]
\centering
\includegraphics[width=0.49\textwidth,height=0.202\textwidth]{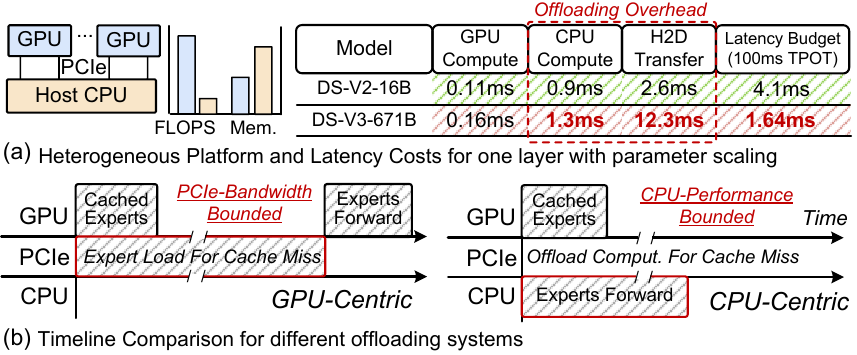}
  \caption{(a) Scaling of MoE offloading overhead against fixed latency constraints on heterogeneous platforms. (b) Timeline illustrating the suboptimal utilization of offloading systems to meet latency demands.}
  \label{fig:offloading}
\end{figure}

However, the parameter scaling of MoEs vastly exceeds GPU VRAM limits. For instance, deploying DeepSeek-V3 requires over 700GB, forcing most weights to be offloaded to host DRAM. To this end, CPU-GPU hybrid inference systems have been developed. Existing solutions typically employ expert-wise offloading and fall into two paradigms (\autoref{fig:offloading} (b)): (1) \textit{GPU-centric approaches} \cite{moe-apex,moe-lightning}, which cache a subset of experts in GPUs and prefetch missing ones via PCIe; and (2) \textit{CPU-centric approaches} \cite{ktransformers,fiddler}, which shift MoE computation to CPUs for absent experts to bypass I/O bottlenecks.

Despite enabling large-scale MoE execution, these systems hit an offloading dilemma. This manifests in two dimensions: latency and efficiency. As parameters grow, the overhead of CPU offloading—including PCIe transfers and CPU computation—scales correspondingly. This growth directly conflicts with the strict and constant latency budget (e.g., 100ms TPOT), leading to latency violations.
Meanwhile, GPUs remain underutilized as the pipeline stalls waiting for CPU computation or I/O, resulting in low overall efficiency.
The root cause lies in a fundamental \textbf{algorithmic-hardware misalignment}. Heterogeneous platforms exhibit pronounced resource asymmetry: GPUs offer high throughput but limited memory, whereas CPUs provide abundant memory capacity but low throughput. Ideally, storage-heavy workloads should reside on CPUs and compute-intensive workloads on GPUs. However, the scaling latency of transferring gigabytes of expert weights between devices forces tight coupling between computation and storage: an expert must be computed where its weights reside. This binding fails to align with hardware, preventing efficient MoE inference.

\begin{figure}[t]
\centering
\includegraphics[width=0.49\textwidth,height=0.25\textwidth]{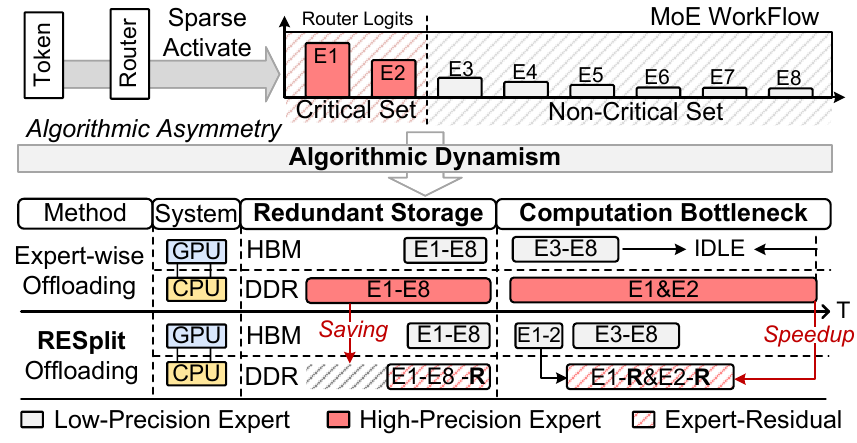}
  \caption{Illustration of algorithmic asymmetry and comparison between expert-wise and RESplit offloading in leveraging asymmetry.}
  \label{fig:fig3n}
\end{figure}

We identify the \textbf{Cross-Asymmetry} in both algorithmic demands and hardware characteristics as the key to break the deadlock. 
On the algorithmic side, MoE routing exhibit skewed importance distributions \cite{expert-pruning1, expert-cache1}(\autoref{fig:fig3n}): a few critical experts demand high-precision storage to maintain model quality, but incurs low floating-point operations (FLOPS) with limited number. Conversely, the remaining non-critical experts tolerate low precision but dominate the overall computation.
 This algorithmic asymmetry naturally aligns with the hardware disparity: CPUs handle the storage-intensive, low-FLOP critical experts, while GPUs excel at the storage-light, high-FLOP non-critical ones.

Exploiting the cross-asymmetry exposes three fundamental challenges in system design. First, expert importance varies across different tokens, layers and inference phases, requiring a runtime arbitration to balance accuracy against latency for \textbf{algorithmic dynamism}.
Second, the dynamic expert importance conflicts with static expert-wise storage, inevitably leading to \textbf{redundant storage} as systems maintain dual copies of high/low-precision across CPUs and GPUs. 
Third, the hardware asymmetry renders the expert-wise scheduling a fundamental \textbf{computation bottleneck}. As even a single expert on CPU becomes the straggler  (\autoref{fig:roofline}), expert-wise offloading fails to achieve efficient utilization.

To address this challenge, we present \textsc{RapidMoE} to exploit Cross-Asymmetry for efficient large-scale MoE inference. 
To tackle algorithmic dynamism, we uncover the inherent multi-level importance divergence in MoE routing and develop a novel \textbf{Unified Multi-Level Importance Arbitration (UMIA)}, which identifies critical experts by integrating \textit{local} routing confidence, \textit{spatial} layer-wise and \textit{temporal} phase-wise importance, achieving an effective trade-off between system latency and accuracy.
To avoid storage and computation dilemmas, We introduce an offloading paradigm shift from expert-level to bit-level through a systematic \textbf{residual-split (RESplit)} framework.
At the \textbf{data representation}, RESplit proposes a residual decomposition to split weights at the bit level, enabling a compact storage and a hardware-aware mapping: memory-intensive residuals $W^R$ are offloaded to CPUs, while quantized components $W^Q$ remain on GPUs.
At the \textbf{routing strategy}, split expert routing is developed to mirror this decomposition in computation. The top-$k$ routed experts are divided into a small set of critical experts (top-$r$) and the remaining top‑($k‑r$) non‑critical experts. Non-critical experts activate only $W^Q$, while critical experts utilize both $W^Q$ and $W^R$ to ensure high-precision computation.
At the \textbf{execution layer}, RESplit expert parallelism provides fine-grained, hardware-aware workload assignments.
Storage-wise, lightweight $W^Q$ is assigned to GPUs while storage-intensive $W^R$ is offloaded to CPUs, avoiding redundant storage.
Computation-wise, high-FLOP non-critical experts are handled by GPUs using $W^Q$ while CPUs refine a minority of critical experts via $W^R$, breaking computation bottlenecks.
Together, these techniques enable \textsc{RapidMoE} to achieve optimal latency and efficiency while maintaining model accuracy. The key innovations are summarized as follows:

\begin{itemize} 
\item We formalize cross-asymmetry as the foundational principle for large-scale MoE inference on heterogeneous platforms: the algorithm-hardware alignment between skewed MoE routing and resource disparity.
\item We introduce UMIA to unify local, spatial, and temporal-wise importance for adaptive discrimination, enabling importance-aware workload partitioning.
\item We propose systematic RESplit methods spanning routing strategy, data representation, and parallelism design, enabling hardware-aware workload partitioning across storage and computation.
\item We implement and evaluate \textsc{RapidMoE} on typical heterogeneous platforms with DeepSeek-671B and Qwen3-235B, achieving up to 3.5$\times$ speedup over SOTA systems.
\end{itemize}

\begin{figure}[t]
\centering
\includegraphics[width=0.5\textwidth,height=0.156\textwidth]{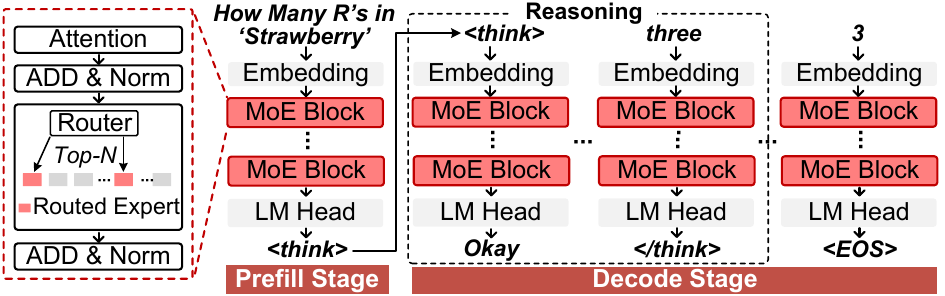}
  \caption{Inference of MoE LLMs with reasoning.}
  \label{fig:moeblock}
\end{figure}

\section{Background}
\subsection{Mixture of Experts and Reasoning Models}
MoE \cite{deepseekmoe,moe1,moe0,gpt4} introduces sparsity by employing multiple expert feed-forward layers and a dynamic gating function to route each input token to the most relevant subset of experts, as shown in \autoref{fig:moeblock}. 
For an input token $x$, the gating function $\mathbf{g}(x)$ computes distribution logits via a linear projection and softmax normalization. Sparsity is enforced by activating the top-$k$ experts with the highest logits. The final output combines the routed experts: $ \sum_{i=1}^{N}\mathbf{g}_i E_i(x)$, 
where $E_i(x)$ denotes the $i$-th expert output, and $\mathbf{g}_i$ is its gating weight. By activating a subset of experts, MoE maintains computation efficiency, for example, DeepSeek-R1 \cite{deepseek-r1} activates 37B of its 671B parameters (k=8). 
The rise of reasoning models further enhances their capabilities \cite{cot2,qwen3}, improving model accuracy by test-time scaling \cite{cot1,o1}.

Despite the benefit, MoE models have far more parameters than dense models of comparable performance. For instance, while achieving similar accuracy, DeepSeek-R1 contains 266B more parameters compared to LLaMA-3.1-405B \cite{llama3}, translating to an extra 532GB of memory requirement. It leads to great costs as all parameters must remain in limited GPU VRAM despite only a small subset being activated. 
Meanwhile, the enhanced performance of reasoning comes at the cost of increased decoding lengths. DeepSeek-R1 achieves 96\% accuracy on MATH-500 but requires an average 9363 output tokens, whereas non-reasoning models average fewer than 2048 tokens on the same benchmark.
This extended decoding time dominates end-to-end latency \cite{seerattention-R}, creating the latency challenge for large-scale MoE serving.

\begin{figure}[t]
\centering
\includegraphics[width=0.49\textwidth,height=0.297\textwidth]{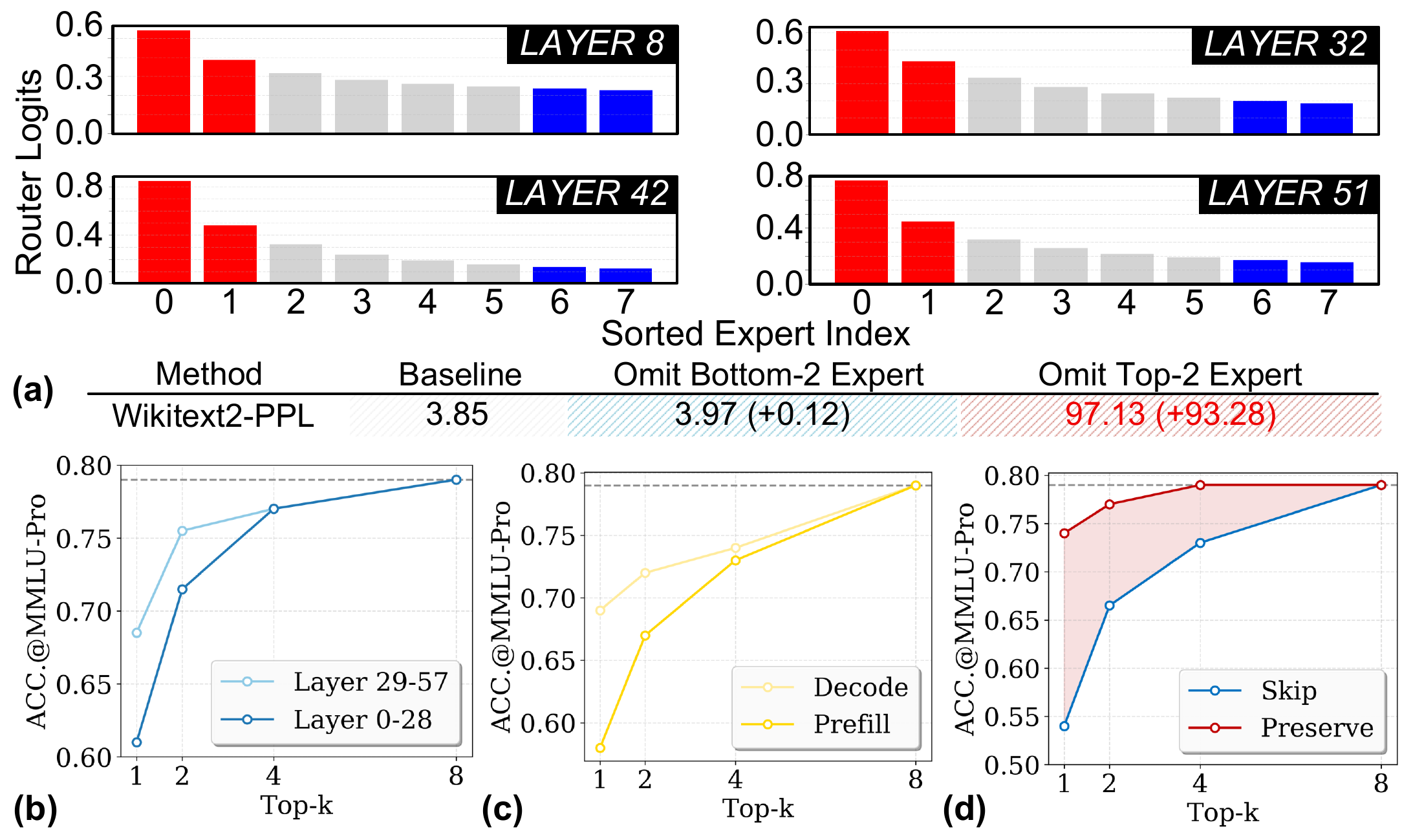}
  \caption{Algorithmic evaluation on DeepSeek-V3: (a) Local divergence highlighted by router distribution and perplexity on Wikitext-2. (b-c) Spatial and temporal divergence evaluated by applying various top-$k$  routed experts. (d) Comparison of expert skipping and expert preserving. } 
  \label{fig:alg_asy}
\end{figure}

\subsection{CPU/GPU Hybrid Inference for MoEs}
To address the deployment challenges, existing systems employ CPU/GPU hybrid inference with expert-wise offloading, which can be categorized into GPU-centric and CPU-centric offloading based on computation device.

Firstly, GPU-centric offloading systems execute MoE layers on GPUs \cite{moe-lightning,moe-apex}. When the required experts are absent, they are transferred from CPU via PCIe connection. Given the limited GPU memory capacity, current works propose Caching and Prefetching.
Caching maintains a subset of experts in GPUs while offloading others to CPUs. The policy of cache is managed by identifying important experts offline \cite{fiddler}, tracking hot experts online and adopting least frequently used for replacement \cite{moe-apex, edgemoe}. Prefetching uses gating logits to predict routed experts in subsequent layers and overlap the expert loading with GPU computation of current layer \cite{moe-apex, moe-infinity}.
Despite addressing capacity concerns, \textbf{these techniques fail to meet the low-latency requirements due to the reliance on CPU-GPU expert loading.} Prefetching attempts to hide loading latency by overlapping, yet the loading time remains on critical path. 
Caching suffers from high miss rates (e.g., 70\% \cite{fiddler}) under limited GPU memory, which in turn forces frequent loading. 
{ Limited PCIe bandwidth results in unacceptable latency. As shown in \autoref{fig:offloading}, transferring 8 experts for a single DeepSeek-V3 layer over PCIe 4.0 takes 12.3 ms—more than \(7\times\) the 1.64 ms latency budget required to achieve a 100 ms TPOT target. }

Secondly, CPU-centric systems offload MoE layer computation to CPUs \cite{llama_cpp,ktransformers,fiddler}. These works design CPU-friendly kernels with SIMD instruction sets \cite{llama_cpp}, and utilize CUDA graphs to reduce inference latency \cite{ktransformers}. 
However, \textbf{this design creates an imbalance workloads, as most workloads are shifted to CPUs while leaving GPUs underutilized.} 
{ As detailed in \autoref{fig:roofline} (c), even when only half the per-layer workload is executed on the CPU (i.e., 4 out of 8 experts), GPU utilization remains limited at only 10--20\%.}
Consequently, it degrades resource utilization and represents a mismatch with hardware disparity, where GPUs offer orders-of-magnitude higher throughput. 
HybriMoE~\cite{hybrimoe} attempts to dynamically schedule partial workloads to GPUs via overlapping CPU-to-GPU weight transfers with computation. 
However, this benefit diminishes for large-scale MoEs; 
the latency of transferring gigabyte-scale experts cannot be amortized by compute overlap (\autoref{fig:offloading}a).

\section{Motivation}
We identify the inherent algorithm-hardware cross-asymmetry as the key to resolving the mismatch.
\subsection{Uncovering Algorithmic Asymmetry}\label{sec3.1}

\textbf{\textit{Observation 1: Multi-Level Importance Divergence.}}

\noindent \textbf{Local Divergence:}
The gate router outputs show skewed distributions where a small subset of experts dominates the outputs (\autoref{fig:alg_asy} (a)). Empirically, omitting the two lowest-scoring experts increases perplexity by merely 0.12, whereas dropping the top-2 experts severely degrades outputs by 93.28. This reveals that \textit{local} importance, captured by routing logits $\mathbf{g}$, is concentrated in a sparse subset with only a few critical experts requiring high-precision. 

\noindent \textbf{Spatial Divergence:}
Router logits only capture input-dependent importance within a single layer, ignoring the varied \textit{spatial} impact across layers. Reducing top-$k$ routing in shallower layers incurs larger quality degradation than in deeper layers as shown in \autoref{fig:alg_asy} (b). This stems from error amplification: errors introduced in early layers propagate through subsequent layers, causing potential explosion \cite{layer-wise}. It suggests that the optimal size of critical set should vary across layers.

\noindent \textbf{Temporal Divergence:}
The nature of LLM inference that operates in prefill and decode phase sequentially introduces a \textit{temporal} divergence \cite{phase-wise}.
Reducing top-$k$ during prefilling degrades accuracy substantially more than during decoding (\autoref{fig:alg_asy} (c)), 
due to the foundational role of prefilling in constructing high-quality KV caches as semantic contexts for subsequent decoding generation.
Thus, system must prioritize higher accuracy during prefilling compared to decoding.

These divergences motivate a unified arbitration framework as detailed in Section \ref{sec:umia}.

\begin{figure}[t]
\centering
\includegraphics[width=0.49\textwidth,height=0.33\textwidth]{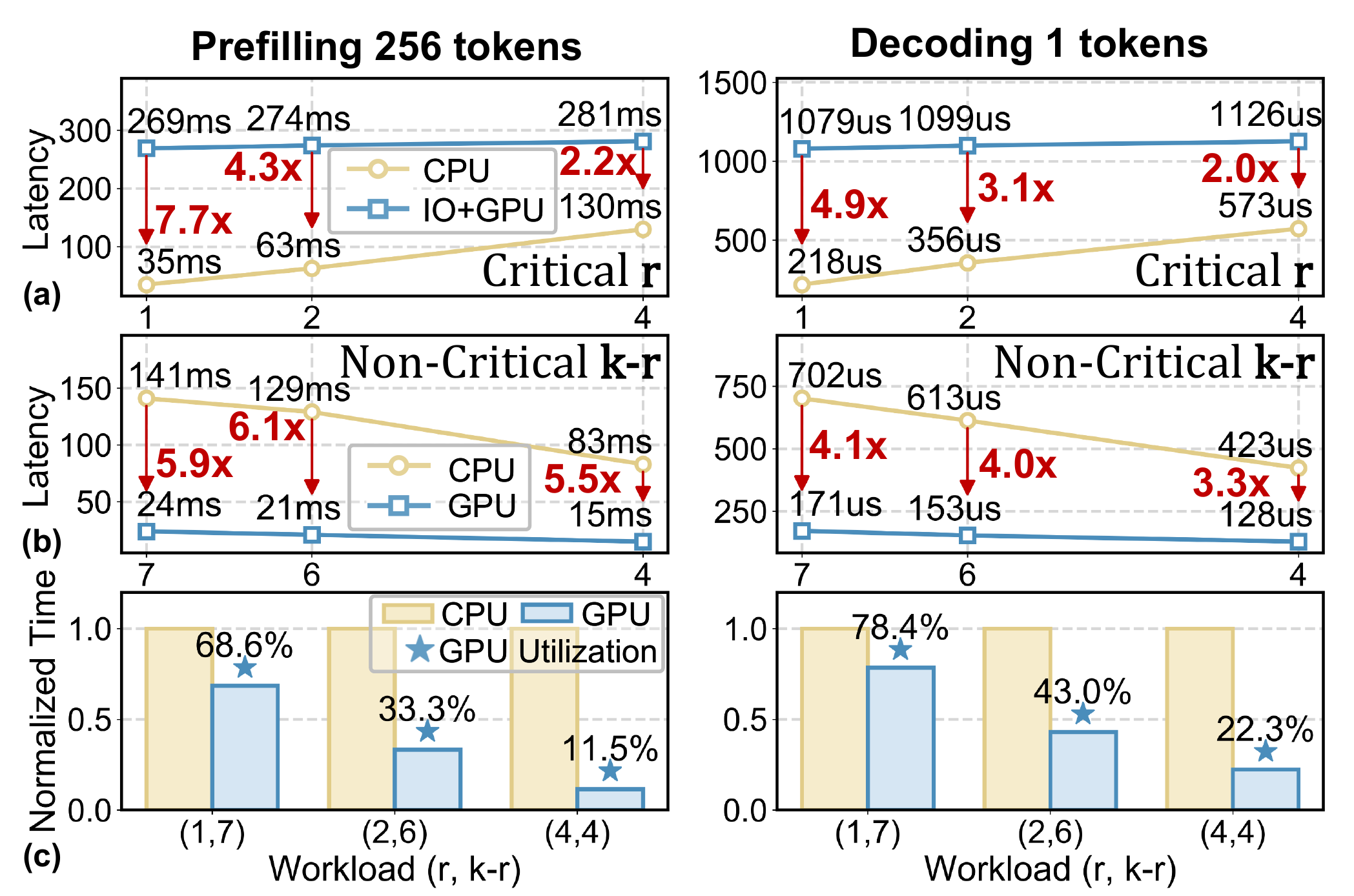}
  \caption{(a-b) Latency of critical (INT4) and non-critical (INT2) experts. Critical experts favor CPUs to bypass PCIe I/O bottlenecks, while non-critical experts maximize throughput on GPUs. (c) Timeline analysis of expert-wise offloading. Even with an ideal mapping (critical on CPUs, non-critical on GPUs), it incurs sub-optimal GPU utilization.}
  \label{fig:roofline}
\end{figure}

\begin{figure*}[t]
\centering
\includegraphics[width=1.0\textwidth,height=0.31\textwidth]{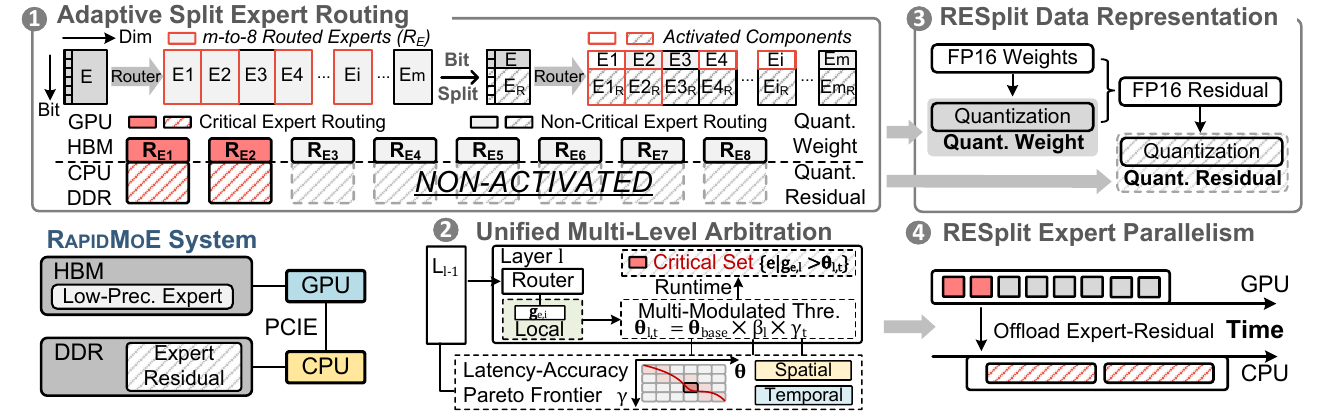}
  \caption{Overview of \textsc{RapidMoE}. UMIA splits routed experts runtime, while RESplit enables efficient storage and execution.}
  \label{fig:overview}
\end{figure*}

\subsection{Leveraging Algorithmic Asymmetry}\label{3.2}

\textbf{\textit{Observation 2: Expert Preserving over Skipping.}}

A naive approach to exploit this asymmetry is \textit{expert skipping}—completely omitting non-critical experts \cite{expert-pruning1,diep}. However, aggressive skipping reduces model generalization capabilities to diverse input workloads. As shown in Figure \ref{fig:alg_asy} (d), skipping the non-critical experts even at $k=4$ causes severe performance loss. In contrast, \emph{expert preserving}, where non-critical experts are preserved in low precision, restores most of the lost generation quality. This suggests that \textbf{while high-score experts are critical for preserving quality, preserving partial information for low-score experts is still essential for robustness.} This motivates our split expert routing to fully realize the theoretical efficiency for \emph{expert preserving} as detailed in Section \ref{3.2}. Specifically, 
routed experts are divided into two sets with complementary resource demands: (1) critical top-$r$ experts, requiring high-precision storage but with less computation workload, and (2) the remaining top-$(k-r)$ experts, which involve more workload but tolerate lower precision.

\subsection{Aligning Cross-Asymmetry}\label{sec3.3}
\noindent \textbf{Cross-Asymmetry Alignment.}  Modern heterogeneous platforms inherently exhibit a compute-memory disparity \cite{flexgen,offload1}. For LLM workloads, GPUs deliver immense computation throughput (e.g., 312 TFLOPS for FP16 on A100) but are bounded by HBM capacity (e.g., 80GB). Conversely, host CPUs offer massive memory capacity (up to 1TB DDR) but possess drastically lower compute throughput. Such disparity demands hardware-aware workload partitioning. Through our profiling at \autoref{fig:roofline}, we identify the cross-asymmetry as the resource demands of MoE experts mirror this hardware disparity:
First, critical experts demand high precision (e.g., INT4) but constitute minimal FLOPs ($r < k-r$). Executing them on the high-capacity CPU bypasses severe PCIe I/O bottlenecks, yielding lower latency. 
Second, non-critical experts tolerate lightweight quantization (e.g., INT2) but dominate the computation workload. Their reduced footprint fits within GPU HBM, fully unleashing the GPU's throughput.
Under parallel execution, latency is bounded by $\max(t_{cpu}, t_{gpu})$. This algorithmic asymmetry naturally compensates for the hardware disparity, effectively closing the execution time gap between $t_{cpu}$ and $t_{gpu}$ and maximizing GPU utilization.

While cross-asymmetry theoretically balances the workload, we observe a fundamental flaw of existing expert-level scheduling paradigm in managing cross-asymmetry:

\noindent \textbf{\textit{Observation 3: Bottlenecks of Expert-Level Scheduling.}}

\noindent \textbf{Redundant Storage.} As the importance varies dynamically, experts cannot be statically pinned to a single precision level (i.e., high or low) for storage. Under the expert-level paradigm, the system is forced to maintain dual copies of weights—both high and low precision—for every expert to cover all routing possibilities, inevitably leading to redundant storage. This memory overhead is non-trivial for large-scale models; for instance, DeepSeek-R1 incurs an additional 170GB memory simply to store 2-bit low-precision replicas.

\noindent \textbf{Computation Bottleneck.} As shown in \autoref{fig:roofline} (c), expert-level scheduling also imposes a fundamental computation bottleneck. Since experts are treated as indivisible atomic units, systems experience GPU bubbles even under an ideal mapping (critical experts on CPUs, non-critical on GPUs) with a minimal CPU workload (e.g., $r=1$).

This reveals that aligning cross-asymmetry requires fine-grained scheduling to overlap execution without redundant storage, which drives the paradigm shift to the RESplit mechanism (Secs. \ref{sec4.4} and \ref{sec4.5}).

\begin{algorithm}[t]
\small
\caption{Split Expert Routing Mechanism}
\label{alg:dp-r}
\KwIn{$x$: input token  \\
$W^H$: high-precision experts weights \\
$W^L$: low-precision experts weights}
\KwOut{$y$: weighted sum }
$\mathbf{g} \leftarrow \text{TopK}(\text{Router($x$)})$ \\
$\mathbf{e}_{c},\mathbf{e}_{n} \leftarrow \text{Unified Multi-Level Importance Arbitration}(\mathbf{g})$ \\
$y \leftarrow 0$ \\
\ForEach{$e \in \mathbf{e}_{c}$} {
    $y \leftarrow y + \mathbf{g}_i\cdot e(x, W^H)$ \tcp{Critical Experts} 
}
\ForEach{$e \in \mathbf{e}_{n}$} {
    $y \leftarrow y + \mathbf{g}_i\cdot e(x, W^L)$ \tcp{Non-Critical Experts}
}
\Return{$y$} 
\end{algorithm}

\section{\textsc{RapidMoE} System}\label{sec4}
\subsection{Design Overview}
We propose \textsc{RapidMoE}, a system for efficient large-scale MoE inference on heterogeneous platforms. 
As illustrated in \autoref{fig:overview}, \textsc{RapidMoE} adopts a co-design approach. 
At the algorithmic level, we propose split expert routing to exploit asymmetry via preserving (Observation 2), which decomposes the top-$k$ routing into top-$r$ critical and top-$(k-r)$ non-critical sets. 
Building upon Observation 1, we introduce UMIA to adjust the critical expert set at runtime, ensuring the accuracy-latency Pareto frontier. 
Informed by Observation 3, we adopt RESplit representation and parallelism to address the storage and computation bottlenecks.
By anchoring low-precision $W^Q$ on GPUs and offloading residuals $W^R$ to CPUs, GPUs deliver computation for non-critical experts utilizing $W^Q$, while CPUs concurrently handle the residual for critical experts with $W^R$.

\begin{figure}[t]
\centering
\includegraphics[width=0.5\textwidth,height=0.36\textwidth]{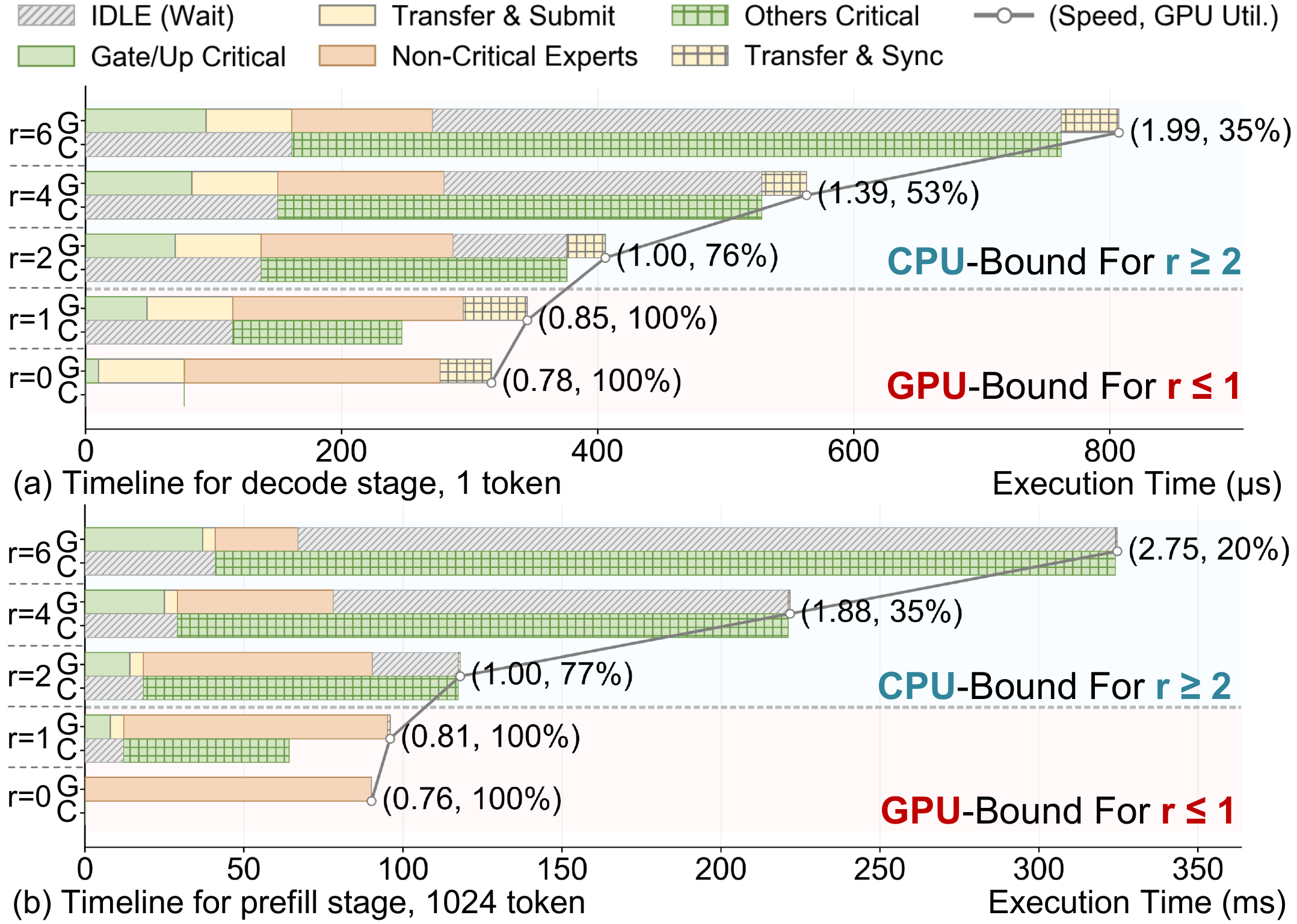}
  \caption{Per-layer RESplit timelines for DeepSeek-V3 at different critical-set sizes $r$ on P1 (\autoref{tab:platform}), with latency normalized to $r=2$: (a) single-token decode and (b) 1024-token prefill. The execution bottleneck shifts at $r=2$.}
  \label{fig:timeline}
\end{figure}

\subsection{Split Expert Routing Mechanism}\label{4.2}
Based on the standard top-$k$ MoE routing \cite{deepseekmoe,qwen3}, we propose the split expert routing mechanism as depicted in Algorithm \ref{alg:dp-r}.
It divides routed experts into critical and non-critical sets based on router logits, selecting the top-$r$ experts as critical. For critical experts, we retrieve high-precision weights $W^H$ to ensure accuracy, while fetching low-precision weights $W^L$ to handle non-critical experts.
Since non-critical experts produce smaller output weights $\mathbf{g}_i$, this design preserves accuracy while reducing computation cost.
Notably, it exploits the inherent router logits in MoE, avoiding any task-specific calibration or retraining.

\noindent \textbf{Asymmetric Latency Scaling.} 
For heterogeneous systems, the latency of an MoE layer can be formulated as:
\begin{equation}
    \label{eq:sys_lat}
    \begin{split}
     \mathbf{T}_{sys} &= t_{sync} +  \max\left(\frac{\text{Load}_{\text{cpu}}}{\text{Perf}_{\text{cpu}}},\frac{\text{Load}_{\text{gpu}}}{\text{Perf}_{\text{gpu}}}\right) \\
    \text{s.t.}\ \  \text{Load}_{\text{cpu}} &\propto r, \quad \text{Load}_{\text{gpu}} \propto k-r
    \end{split}
\end{equation} 
where $t_{sync}$ captures CPU-GPU synchronization overheads. The choice of $r$ represents a trade-off between model accuracy and system latency. To optimally configure $r$, we conduct a latency profiling at \autoref{fig:timeline}, executing the proposed pipeline in Sec.~\ref{sec4.5}. 
The profiling reveals key system behaviors. 
First, the system experiences a bottleneck shift dictated by $\text{Perf}_{\text{gpu}} / \text{Perf}_{\text{cpu}}$. 
The transition point $r^*$, at which CPU and GPU execution times align ($\frac{r}{\text{Perf}_{\text{cpu}}} \approx \frac{k-r}{\text{Perf}_{\text{gpu}}}$), is inherently skewed (e.g., $r^* \in [1, 2]$). Consequently, the system is GPU-bound when $r \le r^*$ and CPU-bound when $r > r^*$.
Second, this shift induces an \textit{asymmetric scaling behavior}: in the CPU-bound region, reducing $r$ lowers latency by alleviating CPU workload, whereas in the GPU-bound regime, further reductions yield diminishing returns.
Third, the synchronization overhead exhibits \textit{phase-dependency}. 
While negligible during the compute-intensive prefill stage, it accounts for $\sim20\%$ of the decode latency, as the short latency of decoding makes a 10-100 $\mu$s synchronization barrier non-negligible. 

Based on these empirical insights, we formulate a simplified relative latency estimator. Let $R(r) = \mathbf{T}_{sys}(r) / \mathbf{T}_{sys}(r^*)$ denote the relative latency for prefill ($R_p$) and decode ($R_d$). We introduce a penalty factor $\eta$ to capture the marginal reduction within the GPU-bound regime, and use $\sigma$ for synchronization overhead:
\begin{equation}\label{eq:lat_est}
    R_{p}(r) = \begin{cases}
    1 - \eta(r^*-r) & r < r^* \\
    \frac{r}{r^*}  & r \ge r^*
    \end{cases}, R_d(r) = \sigma + (1-\sigma)R_p(r)
\end{equation}

\begin{table}[t]
\centering
\small
\caption{Summary of UMIA parameters.}
\begin{tabular}{l l}
\toprule
\textbf{Parameter} & \textbf{Description} \\
\midrule
$\mathbf{g}_e^{(l)}$ & Local importance of expert $e$ in layer $l$ \\
$\beta^{(l)}$ & Spatial importance of layer $l$ \\
$\gamma^{(t)}$ & Temporal importance at phase $t$ \\
$\mathbf{e}_c^{(l,t)}, r^{(l,t)}$ & Critical expert set and its size at layer $l$, phase $t$ \\
$\Theta, \alpha$ & Global base threshold and phase scaling factor \\
$\mathcal{D}, \mathbf{C}, \mathbf{A}$ & Calibration set and candidate sets for $\Theta$ and $\alpha$ \\
$N, K$ & Sizes of $\mathcal{D}$ and candidate sets $\mathbf{C}, \mathbf{A}$ \\
{ $\widehat{T}_{p,d},S_{p,d}$}
& { Estimated prefill/decode latency and their SLOs} \\
\bottomrule
\end{tabular}
\label{tab:umia_params}
\end{table}

\noindent \textbf{System Implication.}  
This asymmetric and phase-dependent relationship reveals that statically constraining the expert number $r$ fails to reflect system latency gain. It directly motivates the system-aware latency constraint formulation in UMIA (Section~\ref{sec:umia}).
In our implementation, we build the estimator with $(\sigma,\eta,r^*)=(0.2,0.15,2)$ based on our profiling. 
While these parameters can be easily calibrated via a one-time micro-benchmark for different platforms, the structural asymmetry of $R(r)$ remains robust to variations across platforms, as it is designed to capture the relative latency scaling trends rather than to provide absolute timing.
This robustness is further validated in the cross-platform evaluation in Sec. ~\ref{sec6} (P1 and P2 in \autoref{tab:platform}).

\subsection{Unified Multi-Level Importance Arbitration}
\label{sec:umia}
{ We introduce Unified Multi-Level Importance Arbitration (UMIA), acting as a training-free, lightweight engine.
 Offline, UMIA calibrates deployment parameters $(\Theta,\alpha,\beta)$ under latency target $S$. At runtime, it uses the MoE router score $\mathbf{g}$ to adaptively determine the optimal critical set size $r$.}

\begin{algorithm}[t]
\small
\caption{Hetero-Aware Frontier Search}
\label{alg:frontier_search}
\KwIn{Calibration set $\mathcal{D}$, SLO constraints $S_p, S_d$, candidate sets $\mathbf{C}, \mathbf{A}$ (ascending order).}
\KwOut{Parameters $(\Theta^*, \alpha^*)$.}

$ \mathcal{F} \leftarrow  \emptyset$\;

\ForEach{prompt $x \in \mathcal{D}$}{
    $\textsc{found} \leftarrow \textbf{false}$\;
    \ForEach{$\Theta \in \mathbf{C}$}{
        \hlstage{gray!10}{%
          \tcp{Stage 1: Prefill forward}
          $r^p \leftarrow \textsc{Prefill}(x;\Theta)$\;
          $\widehat T_p \leftarrow \mathbf{T}_{sys}(r^p)$\;
          \textbf{if} {$\widehat T_p > S_p$}\ \textbf{then} \textbf{continue}\;
        }

        \hlstage{red!10}{%
          \tcp{Stage 2: Decode forward}
          \ForEach{$\alpha \in \mathbf{A}$}{
            \tcp{M Decode Steps}
            $r^d \leftarrow \frac{1}{M}\sum_{i=1}^{M}
            \textsc{Decode}(x;\Theta,\alpha,i)$\;
            $\widehat T_d \leftarrow \mathbf{T}_{sys}(r^d)$\;
            \textbf{if} {$\widehat T_d > S_d$}\ \textbf{then} \textbf{continue}\;
            \tcp{Add Frontier Candidate}
            $\mathcal{F},\textsc{found}\leftarrow \mathcal{F} \cup \{(\Theta,\alpha)\}, \textbf{true}$\;
            \tcp{Early stop: $\mathcal{L}$ increases with $\alpha$}
            \textbf{break}\;
          }
        }
        \tcp{Early stop: $\mathcal{L}$ increases with $\Theta$}
        \textbf{if} {$\textsc{found}$}\ \textbf{then} \textbf{break}\;
    }
}
    \tcp{Evaluate frontier candidates}
\ForEach{$(\Theta,\alpha)\in\mathcal{F}$}{
    $\overline{\mathcal{L}}_p(\Theta)
    \leftarrow
    \frac{1}{|\mathcal{D}|}
    \sum\limits_{x\in\mathcal{D}}
    \textsc{PrefillEval}(x;\Theta)$\;

    $\overline{\mathcal{L}}_d(\Theta,\alpha)
    \leftarrow
    \frac{1}{|\mathcal{D}|M}
    \sum\limits_{x\in\mathcal{D}}\sum_{i=1}^{M}
    \textsc{DecodeEval}(x;\Theta,\alpha,i)$\;
}
\Return{$(\Theta^*, \alpha^*) \leftarrow \text{argmin}_{(\Theta,\alpha)\in \mathcal{F}} (\omega_p\overline{\mathcal{L}}_p + \omega_d\overline{\mathcal{L}}_d)$}\;
\end{algorithm}

\noindent \textbf{Multi-Level Modulation.}
Instead of employing a fixed $r$ or static thresholding, UMIA computes a multi-modulated threshold (MMT) for each layer and execution phase. 
For a given MoE layer $l$ at phase $t$, UMIA defines the MMT $\mathbf{\Theta}^{(l,t)}$ as a multiplicatively separable modulation:
\begin{equation}
\mathbf{\Theta}^{(l,t)} = \underbrace{\Theta}_{\text{Base}} \times \underbrace{\beta^{(l)}}_{\text{Spatial} }\times \underbrace{\gamma^{(t)}}_{\text{Temporal}}
\label{eq:importance-tensor}
\end{equation}
where $\Theta$ is the global base threshold capturing inherent model redundancy, $\beta^{(l)}$ is the layer-wise \textit{spatial} modulation coefficient, and $\gamma_t$ is the phase-wise \textit{temporal} coefficient. { Combined with \textit{local} importance $\mathbf{g}$, UMIA constructs the critical set $\mathbf{e}_c^{(l,t)}(x)$ for inputs $x$ by the gating:}
\begin{equation}
    \mathbf{e}_c^{(l,t)}(x) = \left\{ e \,\big|\, \mathbf{g}_e^{(l)}(x) \geq \Theta^{(l,t)} \right\}, \quad r^{(l,t)} = |\mathbf{e}_c^{(l,t)}|
    \label{eq:critical_set}
\end{equation}
{ The remaining $k - r^{(l,t)}$ experts form the non-critical set. Here, $\mathbf{\Theta}$ controls gate selectivity: a larger $\mathbf{\Theta}$ imposes a stricter gating criterion, admitting fewer critical experts and thus reducing the critical-set size $r$. \autoref{tab:umia_params} summarizes the key UMIA parameters and notation. Specifically, it unifies importance across:}

\begin{itemize}[leftmargin=*]
    \item \textbf{Local Modulator $\mathbf{g}$:} Router logits of MoE layer $l$ naturally capture intra-layer expert importance for given tokens, normalized by softmax such that $\mathbf{g} \in [0,1]$.
    
    \item \textbf{Spatial Modulator $\beta$:} 
    { To quantify layer-wise asymmetry, $\beta$ is derived offline via Kullback-Leibler (KL) divergence as $\beta_l = (\frac{1}{N}\sum_{i=1}^{N}D_{KL}(P_{i} \,\|\, P_{i}^{(l)}))^{-1}$, where $P_{i}$ is the BF16 reference distribution and $P_{i}^{(l)}$ is the quantized counterpart using low-precision experts at layer $l$. A larger $\beta$ therefore denotes a less sensitive layer.}

    \item \textbf{Temporal Modulator $\gamma$:} UMIA models it as a binary-state parameter corresponding to the inference phase:
    \begin{equation}
    \gamma_{t} = 
    \begin{cases} 
    1, & \text{if t is} \text{ Prefill} \\
    \alpha, & \text{if t is} \text{ Decode} 
    \end{cases} 
    \end{equation}
    Based on the analysis at Sec. \ref{sec3.1}, we constrain $\alpha >1 $ to ensure that prefilling utilizes a larger set of $\mathbf{e}_c$ compared to decoding to maintain KV cache quality.
\end{itemize}

\noindent \textbf{Hetero-aware Frontier Search.}
{ Given the router scores $\mathbf{g}$ and KL-calibrated $\beta$, UMIA configures $(\Theta,\alpha)$ via an offline \textit{Hetero-Aware Frontier Search} to balance system latency and model accuracy. 
While prior fixed-sparsity calibration \cite{diep,expert-pruning1} (e.g., globally skipping 75\% of experts) fails to capture the asymmetric scaling and phase-dependency feature as analyzed in Sec. \ref{4.2}, UMIA replaces sparsity constraints with latency constraints $S$: 
\begin{equation}
\begin{split}
\min_{\Theta\in\mathbf{C},\alpha\in\mathbf{A}}\quad
&\overline{\mathcal{L}}(\Theta,\alpha)
=\omega_p\overline{\mathcal{L}}_p(\Theta)
+\omega_d\overline{\mathcal{L}}_d(\Theta,\alpha),\\
\text{s.t.}\quad
&\widehat T_p(\Theta)\leq S_p,\qquad
\widehat T_d(\Theta,\alpha)\leq S_d,
\end{split}
\label{eq:umia_objective}
\end{equation}
$(\Theta,\alpha)$ determine the critical-set sizes $r$ via Eq.~\ref{eq:critical_set}. These sizes are then mapped to estimated execution latency $\widehat T$ using the latency estimator $R(r)\times\mathbf{T}_{sys}(2)$ in Eq.~\ref{eq:lat_est}. $\overline{\mathcal{L}}_{p,d}$ denote the average KL-divergence losses over the calibration set $\mathcal{D}$ for prefill and decode, respectively. The constraints $S$ explicitly bound execution latency (e.g., an SLO equivalent to $\mathbf{T}_{sys}(2)$), rather than impose a fixed number of low-precision experts.

UMIA solves the optimization via a two-stage pipeline in Algorithm \ref{alg:frontier_search}.
Crucially, increasing $(\Theta,\alpha)$ reduces the number of high-precision experts $r$, thereby lowering latency (\autoref{fig:timeline}) and increasing quantization loss, yielding a monotone accuracy--latency trade-off.
UMIA exploits this monotonicity by enumerating candidates in ascending order with early stopping mechanism.
Specifically, for each prompt $x\in\mathcal{D}$, Stage~1 scans
$\mathbf{C}$ in ascending order, skipping SLO-violated candidates before costly decode evaluations.
For each prefill-feasible $\Theta$, Stage~2 similarly finds the first decode-feasible $\alpha$.
Since latency decreases while KL loss increases with $\Theta$ and $\alpha$, the first feasible pair $(\Theta,\alpha)$ minimizes loss within the feasible region, allowing early stopping to terminates the remaining search for the current prompt.
This yields one candidate pair for each $x$, and the resulting pairs collectively form $\mathcal{F}$.
Finally, Alg.~\ref{alg:frontier_search} evaluates each pair in $\mathcal{F}$ over the full calibration set and selects the one minimizing the prompt-averaged weighted loss in Eq.~\ref{eq:umia_objective}.

\begin{algorithm}[t]
\small
\caption{Residual-Split Data Representation}
\label{alg:res_q}
\SetKwProg{Fn}{Function}{}{}
\Fn{RESplit($W$, $b_q$, $b_r$)}{
    \KwIn{$W$: FP16 weight, $b_q$: bit-width, $b_r$: residual bit-width}
    \KwOut{$W^Q$: quantized weight, $W^R$: quantized residual }
    
    $W^Q \leftarrow \text{Quantize}(W, b_q)$
    
    $W^R \leftarrow \text{Quantize}(W - W^Q, b_r)$ \tcp*{Residual} 
    \Return{$(W^Q, W^R)$} 
}
\BlankLine
\Fn{ExpertRESplit($W_{\text{gate}}$, $W_{\text{up}}$, $W_{\text{down}}$, $b_q$, $b_r$)}{
    \KwIn{$\{W_{\text{gate}}, W_{\text{up}}, W_{\text{down}}\}$: Expert weights }
    \KwOut{Quantized expert weights}
    
    \tcp{Gate/Up with RESplit representation}
    $(W^Q_{\text{gate}}, W^R_{\text{gate}}) \leftarrow \text{RESplit}(W_{\text{gate}},b_q,b_r)$ 
    
    $(W^Q_{\text{up}}, W^R_{\text{up}}) \leftarrow \text{RESplit}(W_{\text{up}},b_q,b_r)$ 
    
    \tcp{Down with standard representation}
    $W^L_{\text{down}} \leftarrow \text{Quantize}(W_{\text{down}},b_q)$ 
    
    $W^H_{\text{down}} \leftarrow \text{Quantize}(W_{\text{down}},b_q+b_r)$ 
    
    \Return{$(W^H, W^L)$}
}
\end{algorithm}

\noindent \textbf{Calibration Overhead.} As a \textit{once-for-all} offline optimization, UMIA incurs a modest calibration cost. Profiling $\beta$ requires $\mathcal{O}(LN)$ forward passes (about 30 minutes for $N=16$). 
We configure the search range of $(\Theta,\alpha)$ such that, when $S_p$ and $S_d$ permit similar critical-set sizes ($\Delta r \le 2$), a jointly feasible pair is found within the first three prefill-feasible $\Theta$. 
Frontier construction and final evaluation therefore require $\mathcal{O}(NK)$ and $\mathcal{O}(N^2)$ generations, respectively, for an overall complexity of $\mathcal{O}(NK)$ since $K\gg N$.
With $N,K=16,100$, each generation (1 prefill and 100 decode steps) takes about 10 seconds on a 671B MoE, and the full search completes in about 5 hours---a negligible one-time cost for deployment.
 }

\noindent \textbf{Runtime Arbitration.}
At deployment, UMIA applies the pre-computed $(\Theta,\beta,\alpha)$ via lightweight gating (Eq.~\ref{eq:critical_set}), which can be fused into the MoE inline gating kernel with negligible overhead. 
For multi-token scenarios (i.e., prefill stage and decoding with batch>1), we take the token-averaged $\mathbf{g}$ for UMIA decision-making.
The MMT dynamically partitions workload between CPUs and GPUs, ensuring the system operates at a favorable accuracy--latency frontier.

\subsection{Residual-Split Data Representation}\label{sec4.4}
As revealed in Observation 3, the varying expert importance introduces notable memory overhead by dual-weight storage. For instance, DeepSeek-R1 incurs an additional 170GB memory demand even with 2-bit $W^L$. Therefore, we implement RESplit data representation to eliminate redundant storage.

\noindent \textbf{Residual-Split Representation.} We employ a two-stage split representation as detailed in Algorithm \ref{alg:res_q}.
First, weights are quantized to low precision as $W^Q$. The residual $W-W^Q$ is then quantized as $W^R$. This two-stage approach preserves weight information and reduces quantization error through residual compensation. High and low-precision weights can be represented by $W^H = W^Q + W^R, W^L = W^Q$. While avoiding dual-copy storage, it also enables efficient parallel computation in CPU-GPU systems through linear decomposability: $XW^H = XW^Q(\text{GPU}) + XW^R (\text{CPU})$.

\noindent \textbf{GLU Architecture Adoption.} For linear layers, residual quantization enables parallel computation through the decomposition $XW^Q + XW^R$. However, the implementation becomes challenging due to gated linear unit (GLU) structure. 
The GLU computation $\text{O} = W_{\text{down}}(\text{SiLU}(W_{\text{gate}} x \odot W_{\text{up}} x))$ introduces nonlinearity that disrupts parallelization for \textit{down\_proj} layer.
Specifically, when $W^Q$ and $W^R$ are stored separately, \textit{gate\_proj} and \textit{up\_proj} maintain linearity and can be decomposed as $XW^Q_{\text{gate/up}} + XW^R_{\text{gate/up}}$. However, the non-linear Swish function before \textit{down\_proj} requires full intermediate results, forcing activation gathering and synchronization after \textit{gate\_proj} and \textit{up\_proj}.
Such frequent CPU-GPU synchronization fragments kernels, increases launch overhead and undermines CUDA graph optimization.
To balance computation and storage efficiency, we apply residual quantization only to \textit{gate\_proj} and \textit{up\_proj}, while employing dual-quantization for \textit{down\_proj}. 
\autoref{fig:res} shows an execution example of critical expert forwarding, with CPU-GPU synchronization only at the start and end of the operator.
The detailed computation flow is discussed in Sec. \ref{sec4.5}.

\subsection{Residual-Split Expert Parallelism}\label{sec4.5}
Expert-wise offloading that assigns high-precision experts to GPUs and low-precision ones to CPUs inevitably leads to GPU bubbles as demonstrated in \autoref{fig:roofline}.
We therefore develop RESplit parallelism to fully exploit GPU throughput.

\noindent \textbf{Memory Allocation.} \textsc{RapidMoE} employs a compact, structured memory allocation where expert weights are stored sequentially as low-precision $W^Q$ followed by residuals $W^R$.
At initialization, $W^Q$ and $W^R$ are stored in CPU memory, with $W^Q$ subsequently loaded onto GPUs.
This design offers two key advantages: (1) tokens can be processed through low-precision $W^Q$ on GPUs; and (2) CPUs only need to access the lightweight residuals $W^R$ to refine a small set of critical experts. This approach, termed Expert-Residual Offloading, is elaborated in the next paragraph.

\begin{figure}[t]
\centering
\includegraphics[width=0.5\textwidth,height=0.24\textwidth]{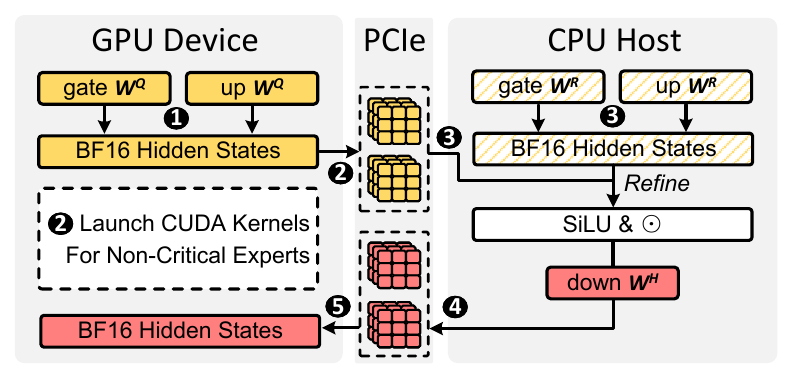}
  \caption{Execution example for critical experts with residual-offloading.}
  \label{fig:res}
\end{figure}

\noindent \textbf{Expert-Residual Offloading.} 
Experts follow a Swish-gated linear unit structure: $\text{O} = W_{\text{down}}(\text{SiLU}(W_{\text{gate}} x \odot W_{\text{up}} x))$. 
We decompose the pre-Swish computations as $h_{\text{up/gate}} = W_{\text{up/gate}}^{Q}x + W_{\text{up/gate}}^{R}x$, where $W^Q$ and $W^R$ denote low-precision weights and quantization residuals, respectively.
With the split layout in \textsc{RapidMoE}, this design naturally enables heterogeneous computation:
\begin{equation}
\underbrace{W_{\text{down}}}_{\text{CPU}} \cdot \left( \text{SiLU}\left( \underbrace{W^Q_{\text{gate}} x}_{\text{GPU}} + \underbrace{W^R_{\text{gate}} x}_{\text{CPU}} \right) \odot \left( \underbrace{W^Q_{\text{up}} x}_{\text{GPU}} + \underbrace{W^R_{\text{up}} x}_{\text{CPU}} \right) \right)
\end{equation}
where GPUs handle the $W_{\text{gate/up}}^Qx$ terms, and CPUs process the residual $W_{\text{gate/up}}^R x$ terms. It further alleviates CPU bottlenecks while preserving computation accuracy.

\noindent \textbf{Computation Flow.} 
{ \autoref{fig:timeline} shows the detailed RESplit timeline, with corresponding steps marked in \autoref{fig:res}.
For each input token and its routed experts, GPUs first compute the \textit{gate\_proj} and \textit{up\_proj} of critical experts (\textbf{Gate/Up Critical}, \stepnum{1}).
After transferring these activations to CPUs (\textbf{Transfer\&Submit}), CPUs launch computation threads while GPUs concurrently process non-critical experts using low-precision $W^Q$ (\textbf{Non-Critical Experts}, \stepnum{2}).
In the residual forward pass (\textbf{Others Critical}), CPUs access $W^R$ for \textit{gate\_proj} and \textit{up\_proj}, compute residual hidden states, and merge them with GPU-computed states (\stepnum{3}). CPU outputs are then generated through Swish activation and high-precision \textit{down\_proj} (\stepnum{4}).
The system perform synchronization to derive final outputs from GPUs and CPUs (\textbf{Transfer\&Sync}, \stepnum{5}).
}

This parallelism benefits both decode and prefill phases by fine-grained scheduling that reserves only critical expert residuals for CPUs. 
Compared with the naive scheduling pipeline in \autoref{fig:roofline}, it achieves higher GPU utilization across various settings of $r$, improving from 33\% to 77\% at $r=2$, and from 69\% to 100\% at $r=1$. 

\section{Implementation Details}
\label{sec5}
We build \textsc{RapidMoE} on top of KTransformers~\cite{ktransformers} framework, which provides a flexible infrastructure for kernel injection and hybrid CPU-GPU execution. Residual-split parallelism is implemented using fine-grained CUDA stream scheduling, enabling the parallel execution across CPUs, GPUs and PCIe transfers.
For MoE operations, we design specialized C++ kernels for CPU expert-residual forwarding and flexible CUDA kernels for GPU residual-offloading. Both are capable of being capture by CUDA graphs to eliminate kernel launch overhead.
For non-MoE operations, we utilize the MARLIN kernels \cite{marlin} for linear layers and the customized flashinfer backend \cite{flashinfer,ktransformers} for attention.

\noindent \textbf{Quantization Framework.}
We adopt the GPT-generated Unified Format (GGUF) \cite{gguf} for cross-platform compatibility. 
For resource-limited deployment, weight-only W4A16 quantization (e.g., AWQ \cite{awq}, GPTQ \cite{gptq}, \textsf{Q4\_K} 
\cite{llama_cpp}) has become the de facto standard, offering effective compression with minimal accuracy loss.
Following established approaches, we employ \textsf{Q4\_K} method as baseline and develop our strategy.
Specifically, low-precision $W^Q$ are quantized in \textsf{IQ1\_M\_R4} format, an average 1.75-bit method implemented via codebook compression \cite{ik_llama_cpp,quip_sharp}. For residuals $W^R$, we employ \textsf{Q2\_K\_R4} format with an average 2.5 bits.
Our experiments show that the combination of \textsf{IQ1\_M\_R4} and \textsf{Q2\_K\_R4} for $W^Q+W^R$ (total 4.25 bits) achieves accuracy comparable to 4.5-bit \textsf{Q4\_K} and AWQ. Non-MoE weights are in 8-bit \textsf{Q8} as they account for a negligible fraction of total weights.

\noindent \textbf{Memory-Adaptive Deployment.}
\textsc{RapidMoE} offers adaptability for diverse platforms. At the system level, we integrate caching for low-precision $W^Q$. When VRAM is insufficient to accommodate $W^Q$, the system executes the standard RESplit path for cached $W^Q$ on GPUs, and falls back to a CPU-centric path upon a cache miss—computing $W^Q$ for non-critical experts and $W^Q+W^R$ for critical ones on CPUs. 
At the algorithmic level, the VRAM footprint is highly elastic; it can be statically tailored to specific hardware budgets by adjusting model scales or employing different quantization for $W^Q$. 
Together, this dual-level adaptability ensures \textsc{RapidMoE} maintains flexibility under arbitrary constraints.

\begin{table}[t]
\footnotesize
\centering
\caption{Configuration of evaluated large-scale MoE models.}
\label{tab:moe}
\begin{tabular}{@{}l @{\hspace{0.5em}}|@{\hspace{0.5em}} c @{\hspace{0.5em}}|@{\hspace{0.5em}} c @{}}
\toprule
MoE LLMs & \makecell{\textbf{DeepSeek-V3/R1}} & \makecell{\textbf{Qwen3-235B-A22B}} \\
\midrule
\#Total Parameters & 671B & 235B \\
\#Activated Parameters & 37B & 22B \\
\#Expert Nums & 256 & 128 \\
\#Routed Expert/top-k & 8 & 8 \\
\#Shared Expert & 1 & 0 \\
\bottomrule
\end{tabular}
\end{table}

\noindent \textbf{UMIA Configuration \& Frontier Search.} 
We sample a calibration set $\mathcal{D}$ ($N=16$) from WikiText-2 \cite{wikitext103}. For frontier search, the candidate sets $\mathbf{C} \in (0, 1)$ and $\mathbf{A} \in (1, 1.5)$ are sampled with a resolution of $K=100$. To prioritize foundational prefill quality, the joint objective weights are set to $\omega_p = 1.5$ and $\omega_d = 1.0$. The latency constraints are set as $S = \mathbf{T}_{sys}(2)$.

\section{Evaluation}\label{sec6}
\subsection{Evaluation Setup}
\noindent \textbf{Models.}
We evaluate \textsc{RapidMoE} on SOTA large-scale MoE models with representative architectures that are mainstream and well-received in the community: DeepSeek-V3-671B \cite{deepseek-v3}, DeepSeek-R1-671B \cite{deepseek-r1}, Qwen3-235B-A22B \cite{qwen3}.
As shown in \autoref{tab:moe}, DeepSeek-V3/R1 employs 256 experts per layer, with each token activating 8 routed experts plus 1 shared expert. Qwen3 adopts a more compact 128 expert configuration, routing each token to 8 experts without shared expert.

\begin{table}[t]
\centering
\caption{Hardware configurations.}
\label{tab:platform}
\footnotesize  
\begin{tabular}{lcc}
\toprule
\textbf{Platform} & \textbf{P1} & \textbf{P2} \\
\midrule
GPU & A800-80G & RTX 4090-24G   \\
CPU (Intel Xeon) &  Gold 6326 (64 core)& Platinum 8375C (128 cores) \\
Memory & 512GB DDR4-3200 & 512GB DDR4-3200  \\
DDR Bandwidth & 220 GB/s & 220 GB/s \\
PCIe & 4.0 x 16 & 4.0 x 16 \\
\bottomrule
\end{tabular}
\end{table}

\begin{table}[t]
\centering
\caption{Evaluation setups.}
\label{tab:configs}
\footnotesize  
\begin{tabular}{lccc}
\toprule
\textbf{Configuration} & \textbf{Model} & \textbf{Platform} &\textbf{GPU Usage} \\
\midrule
\makecell{C11} & DeepSeek-R1 & P1 & A800$\times$2 (160GB)     \\
\makecell{C12} & DeepSeek-V3 & P1 & A800$\times$2 (160GB)    \\
\makecell{C13} & Qwen3-235B-A22B & P1 & A800$\times$1 (80GB) \\
\cmidrule(r){1-4}
\makecell{C21} & DeepSeek-R1 & P2 & 4090$\times$7 (168GB)\\
\bottomrule
\end{tabular}
\end{table}

{
\noindent \textbf{Hardware and Placement.} 
\autoref{tab:platform} details evaluation platforms: (1) \textbf{P1} featuring A800-80G GPUs with 512GB CPU DDR4 memory, and (2) \textbf{P2} utilizing consumer-grade RTX 4090 GPUs with 512GB DDR4-3200.
Both platforms use PCIe for GPU interconnect, with P1 and P2 comprising one and two NUMA domains, respectively.
To evaluate efficiency in memory-constrained environments, we test multiple configurations listed in \autoref{tab:configs}.
For DeepSeek-R1 and V3 that require 700GB VRAM, we allocate 160GB GPU memory that sufficient for storing all low-precision experts. The smaller Qwen3 operates under an 80GB memory constraint.
For multi-GPU execution, model placement is partitioned by layer: each GPU stores and executes a disjoint subset of layers. The CPU-resident weights corresponding to each GPU's layer subset are pinned in the NUMA domain local to that GPU, minimizing cross-NUMA memory traffic.
}

{
\noindent \textbf{Metrics.}
We evaluate system performance through prefill and decode throughput in tokens/s. For serving, we report time to first token (TTFT) for prefilling, and time per output token (TPOT) for decoding, including p50 and p99 latencies. Batch size denotes the number of requests co-executed in a model forward pass, whereas concurrency denotes the maximum number of in-flight serving requests.
}

\noindent \textbf{Baselines.}
We consider the following systems as baselines:
\begin{itemize}[wide=0pt]
    \item \texttt{MoE-APEX} \cite{moe-apex}, a GPU-centric system that employs adaptive precision expert offloading to reduce inference latency.
    \item \texttt{HybriMoE} \cite{hybrimoe}, a CPU-centric system that utilizes dynamic scheduling and score-based caching to enhance resource utilization and workload balance.
    \item \texttt{KTransformers} \cite{ktransformers}, one of the most popular systems for MoE inference on memory-constrained environments with optimized CPU kernels and scheduling strategy.
    \item \texttt{llama.cpp} \cite{llama_cpp} version b6030, a lightweight C/C++ implementation for efficient LLM inference. It can statically map model layers to CPU or GPU for execution.
\end{itemize}

{ 
For a fair comparison, baselines deploy similarly sized \textsf{Q4\_K\_M} models from Unsloth \cite{unsloth}.
Specifically for MoE-APEX that supports adaptive precision, we align its precision as INT4 (\textsf{Q4\_K\_M}) for high-precision and INT2 (\textsf{IQ1\_M\_R4}) for low-precision. 
We align CPU and GPU GEMM/GEMV kernel for all methods from \cite{llama_cpp}, sharing identical kernels for the same quantization format and using format-specific kernels only for different dequantization requirements.

We configure all baselines with comparable memory budgets. As \texttt{llama.cpp} only supports layer-wise offloading, 22 of DeepSeek-R1’s 61 layers are loaded to the GPU, consuming 149GB of the available 160GB GPU memory.
For others that employ expert-wise offloading, 40\% of experts per layer are cached on GPUs while the remaining on CPUs.
Specifically for \texttt{MoE-APEX}, we evaluate its two computing modes as variants: \texttt{MoE-APEX-G} (GPU-Centric Mode), which loads cache-missed experts via PCIe, and \texttt{MoE-APEX-C} (GPU-CPU Cooperative Mode), which executes missed experts on CPUs. 

}

\begin{table}[t]
\centering
\caption{Accuracy evaluation under batch size 1.}
\footnotesize
\label{tab:acc}
\begin{tabular}{@{}l 
@{\hspace{0.5em}}|@{\hspace{0.5em}} c
@{\hspace{0.5em}}|@{\hspace{0.5em}} c
@{\hspace{0.5em}}|@{\hspace{0.5em}} c 
@{\hspace{0.5em}}|@{\hspace{0.5em}} c @{\hspace{0.5em}}|@{\hspace{0.5em}} c 
@{\hspace{0.5em}}|@{\hspace{0.5em}} c  @{}}
\toprule
 \multicolumn{2}{c}{\multirow{2}{*}{\textbf{Benchmark}}}\tnote{1)} &  \multicolumn{1}{|c|}{\tiny GENERAL}& \multicolumn{2}{c|}{\tiny MATH} & \multicolumn{2}{c}{\tiny CODE} \\
\multicolumn{2}{l}{\multirow{2}{*}{ }} & \multicolumn{1}{|c|}{\tiny MMLU-Pro} & \multicolumn{1}{c}{\tiny MATH-500} &\multicolumn{1}{c|}{\tiny AIME-24} & \multicolumn{1}{c}{\tiny HumanEval} & \multicolumn{1}{c}{\tiny EvalPlus}  \\
\midrule
\rowcolor{gray!20} \multicolumn{7}{c}{\textit{DeepSeek-R1, Reasoning Model}} \\
\multicolumn{1}{c}{\textsf{Q4\_K\_M}} &\multicolumn{1}{c|}{380GB} & \multicolumn{1}{c|}{83.9} & \multicolumn{1}{c}{95.6} & \multicolumn{1}{c|}{73.3} & \multicolumn{1}{c}{98.2} & \multicolumn{1}{c}{-}\\ 
\multicolumn{1}{c}{\text{\textsc{RapidMoE}}}&\multicolumn{1}{c|}{390GB} & \multicolumn{1}{c|}{\text{83.8}} & \multicolumn{1}{c}{\text{96.2}} & \multicolumn{1}{c|}{\text{73.3}}& \multicolumn{1}{c}{\text{98.2}} & \multicolumn{1}{c}{-} \\
\midrule
\rowcolor{gray!20} \multicolumn{7}{c}{\textit{Qwen3-235B-A22B, Reasoning Model}} \\
\multicolumn{1}{c}{\textsf{Q4\_K\_M}} &\multicolumn{1}{c|}{133GB} & \multicolumn{1}{c|}{81.0} & \multicolumn{1}{c}{96.4} & \multicolumn{1}{c|}{70.0} & \multicolumn{1}{c}{96.3} & \multicolumn{1}{c}{-} \\ 
\multicolumn{1}{c}{\text{\textsc{RapidMoE}}}&\multicolumn{1}{c|}{140GB}& \multicolumn{1}{c|}{\text{80.5}} & \multicolumn{1}{c}{\text{96.0}} & \multicolumn{1}{c|}{\text{73.3}}& \multicolumn{1}{c}{\text{97.0}} & \multicolumn{1}{c}{-}\\
\midrule
\rowcolor{gray!20} \multicolumn{7}{c}{\textit{DeepSeek-V3, Non-Reasoning Model}} \\
\multicolumn{1}{c}{\textsf{Q4\_K\_M}} &\multicolumn{1}{c|}{380GB}& \multicolumn{1}{c|}{78.0} & \multicolumn{1}{c}{94.8} & \multicolumn{1}{c|}{56.7} & \multicolumn{1}{c}{93.3} & \multicolumn{1}{c}{82.5}\\ 
\multicolumn{1}{c}{\text{\textsc{RapidMoE}}}&\multicolumn{1}{c|}{390GB} & \multicolumn{1}{c|}{\text{78.2}} & \multicolumn{1}{c}{95.2} & \multicolumn{1}{c|}{56.7}& \multicolumn{1}{c}{\text{92.7}} & \multicolumn{1}{c}{\text{82.6}} \\
\bottomrule
\end{tabular}
\begin{tablenotes}
\footnotesize
\item[1)] (1) Validation (except EvalPlus) uses Evalscope \cite{evalscope}.
\item[2)] (2) Reasoning models are excluded from EvalPlus as the framework lacks support for their evaluation.
\end{tablenotes}
\end{table}

{
\noindent \textbf{Datasets and Workloads.}
We construct the performance workloads from ShareGPT. Prefill experiments vary input length from 256 to 4096 tokens, while decode experiments use 512-token input prompts by default. 
For serving workload, we sample 1000 ShareGPT requests with an average input length of 715 tokens and cap the output length at 512 tokens. For accuracy evaluation, we benchmark on: (1) MMLU-Pro \cite{mmlu-pro} (1000 randomly sampled items) for general knowledge and reasoning evaluation, (2) MATH-500 \cite{math500} and AIME-24 \cite{aime-24} for complex mathematical problem-solving, and (3) HumanEval \cite{humaneval} and EvalPlus \cite{evalplus} for code generation. 
}

\subsection{End-to-end Accuracy}
As shown in \autoref{tab:acc}, \textsc{RapidMoE} matches or exceeds \textsf{Q4} accuracy. It maintains accuracy with -0.1\% in general knowledge, +0.7\% in mathematical reasoning, and +0.1\% in code generation. This advantage holds for both reasoning models, which generate longer outputs via test-time scaling, and non-reasoning models with shorter outputs.

\begin{table}[t]
\centering
\small
\setlength{\tabcolsep}{3pt}
\caption{Accuracy at batch size 4, with changes relative
to batch size 1.}
\label{tab:batched_decode_acc}
\resizebox{\columnwidth}{!}{%
\begin{tabular}{lcccc}
\toprule
\textbf{Model}
& \textbf{MMLU-Pro}
& \textbf{MATH-500}
& \textbf{AIME-24}
& \textbf{HumanEval} \\
\midrule
DeepSeek-R1
& 83.7 ($-0.1$)
& 96.0 ($-0.2$)
& 73.3 ($0.0$)
& 98.2 ($0.0$) \\
DeepSeek-V3
& 78.0 ($-0.2$)
& 94.8 ($-0.4$)
& 56.7 ($0.0$)
& 92.3 ($-0.4$) \\
\bottomrule
\end{tabular}%
}
\end{table}

\begin{table}[t]
\centering
\small
\setlength{\tabcolsep}{5pt}
\caption{ Calibration sensitivity of DeepSeek-V3 across: $\{\text{ShareGPT},\text{WikiText-2},\text{MMLU-Pro}\}\times\{8,16,32\}$ prompts.}
\label{tab:calibration_sensitivity}
\resizebox{\columnwidth}{!}{%
\begin{tabular}{
>{}c
>{}c
>{}c}
\toprule
$\beta$ Variation
& Max. Parameter Difference
& Max. Acc. Difference \\
\midrule
$4.1$--$7.2\%$
& \makecell{$|\Delta\Theta|,|\Delta\alpha|=0.03,0.025$\\
($\leq 5/100$ grid positions)}
& \makecell{$0.0\%$@AIME-24\\$-0.4\%$@MMLU-Pro} \\
\bottomrule
\end{tabular}
}
\end{table}

\begin{figure}[t]
\centering
\includegraphics[width=0.5\textwidth,height=0.26\textwidth]{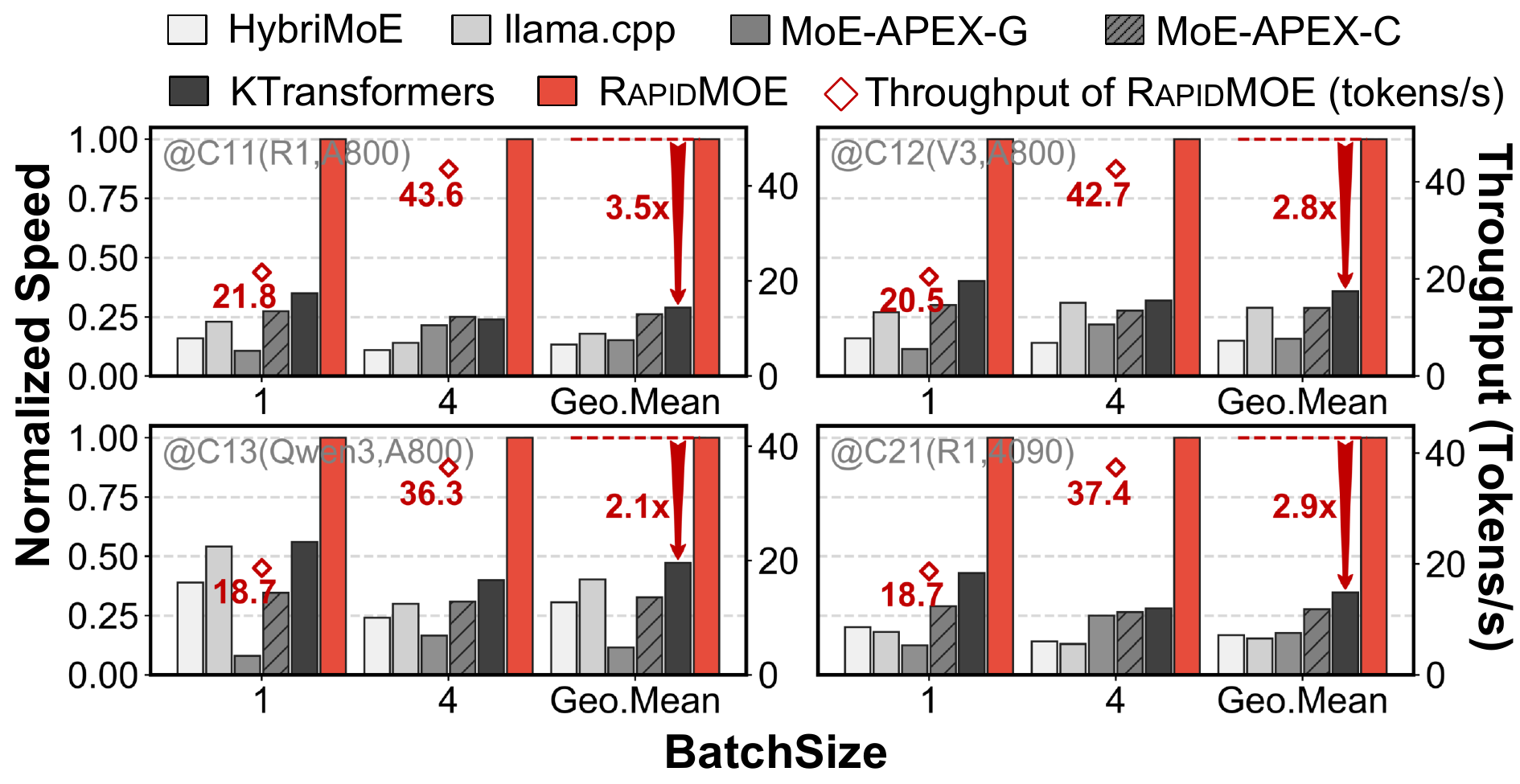}
  \caption{Evaluation of decoding speed on DeepSeek-R1/V3 and Qwen3 models compared in different scenarios.}
  \label{fig:decode}
\end{figure}

{
\noindent \textbf{Robustness.}
RapidMoE exhibits robustness to token aggregation and calibration variations. 
For prefill and batched decoding, UMIA averages router scores across tokens before critical-set selection. \autoref{tab:batched_decode_acc} evaluates accuracy with batch size of 4, where token-averaged routing is applied during prefill and decode. 
For different calibration datasets, \autoref{tab:calibration_sensitivity} shows stable accuracy, with normalized $\beta$ varies by 4.1--7.2\%, while the maximum differences in $\Theta,\alpha$ vary by at most 5 of 100 grid positions. 
 Such robustness is facilitated by UMIA’s mapping of continuous parameters and router scores to the discrete size $r$, allowing small perturbations to be absorbed.

}

\subsection{End-to-end Efficiency}
\subsubsection{Decode Stages}\label{sec6.3.1}
\autoref{fig:decode} presents the decoding performance of \textsc{RapidMoE} across four configurations (\autoref{tab:configs}). \textsc{RapidMoE} consistently dominates all scenarios. On \textbf{P1} A800 platforms, it achieves averaged speedups of 3.5$\times$, 2.8$\times$, and 2.1$\times$ for R1, V3, and Qwen3, respectively, over the SOTA KTransformers. Crucially, these gains extend to \textbf{P2}, yielding up to 2.9$\times$ speedup with different CPU-GPU setting, demonstrating robust cross-platform efficiency.

\begin{figure}[t]
\centering
\includegraphics[width=0.5\textwidth,height=0.33\textwidth]{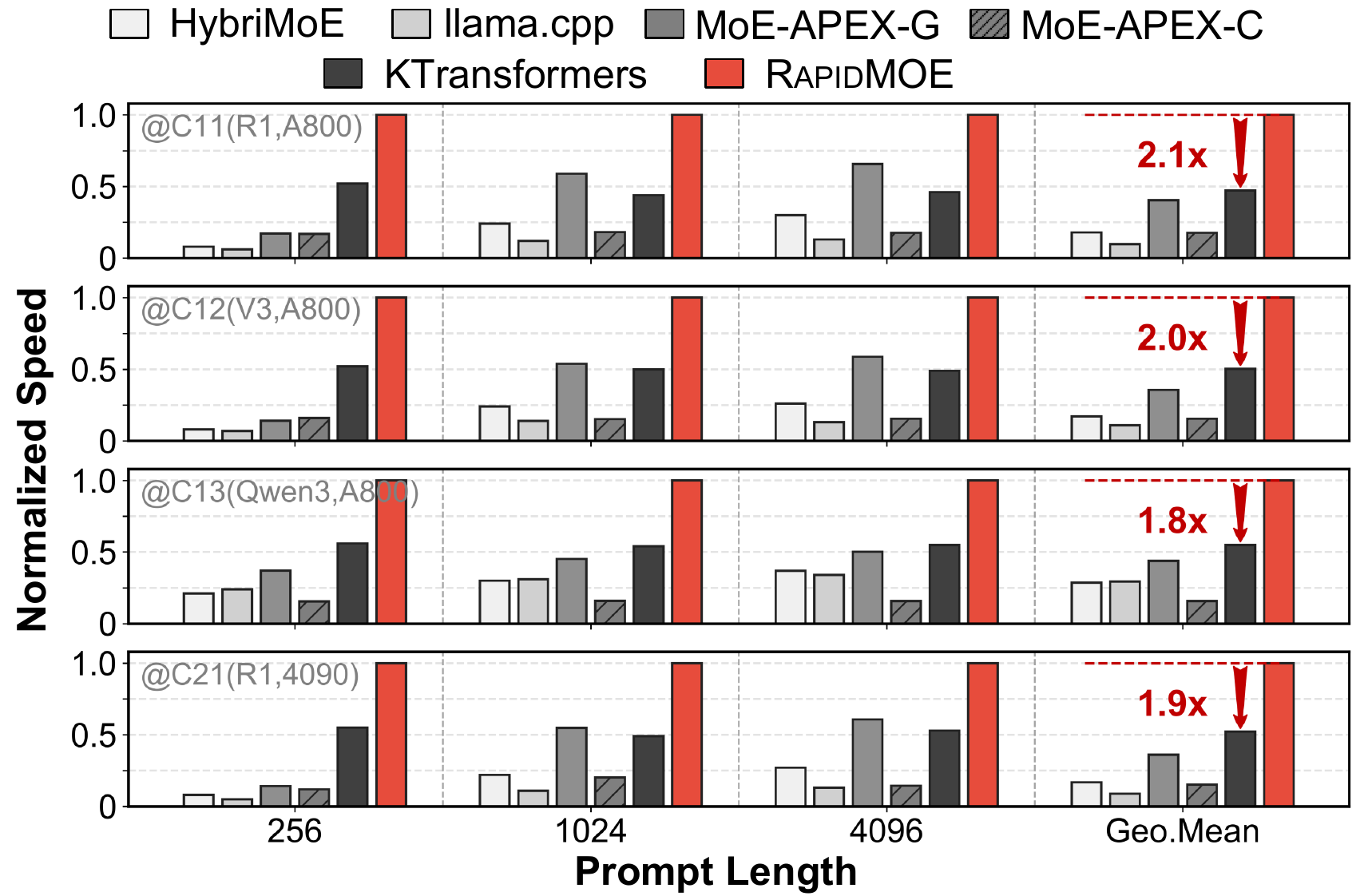}
  \caption{Evaluation of prefilling speed on DeepSeek-R1/V3 and Qwen3 models compared in different scenarios.}
  \label{fig:prefill}
\end{figure}

Our evaluation exposes the fundamental bottlenecks of prior \textit{expert-wise} offloading paradigms. GPU-centric methods (e.g., MoE-APEX-G) suffer the slowest decoding speeds due to weight transfer latency. Conversely, CPU-centric approaches (KTrans., HybriMoE) bypass PCIe bottlenecks but leave GPUs idle. 
Notably, while MoE-APEX attempts adaptive precision, its reliance on purely \textit{local} gating logits fails to exploit multi-level divergence. 
Based on the profiling, it leads to overly preserving of 3-4 high-precision experts per layer for DeepSeek-V3, resulting in either IO bottlenecks (APEX-G) or CPU compute bottlenecks (APEX-C). 
In contrast, \textsc{RapidMoE} benefits from its cross-stack co-design. Algorithmically, UMIA maintains the accuracy-latency Pareto frontier by dynamically capturing a minimal critical set. System-wise, fine-grained residual offloading optimally distributes the workload and reduces GPU bubbles.
Furthermore, \textsc{RapidMoE} exhibits scalability with increasing batch sizes (\autoref{tab:throughput}), benefiting from improved computation efficiency in GEMM and attention kernels at larger batches. {
\textit{Long-Output} in \autoref{tab:throughput} further evaluates AIME-24 reasoning workloads for reasoning models R1 and Qwen3, showing that \textsc{RapidMoE}'s decode throughput persists for long-output generation.
}

\subsubsection{Prefill Stages}\label{sec6.3.2}
\autoref{fig:prefill} presents the prefill performance. \textsc{RapidMoE} improves prefilling throughput across varying sequence lengths, achieving 138.9 tokens/s on DeepSeek-V3 and exhibiting an average 2 $\times$ higher throughput than baselines. Notably, this speedup is less substantial than in the decode stage. This discrepancy arises as residual offloading transfers partial weight access cost to GPUs, leaving CPUs to compute only the lower-bitwidth $W_R$. This significantly reduces memory access overhead, yielding a greater advantage during the memory-bound decode phase.
{
For long-context inputs (>4K), \textsc{RapidMoE} employs layer-wise overlap, transferring next-layer weights while GPUs compute the current layer. Under C11/C12, each H2D transfer costs 136 ms, while per-layer computation for an 8K input takes 116 ms, hiding most of the transfer latency. Longer inputs further increase the overlap. Across 8K, 12K, and 16K inputs, \textsc{RapidMoE} achieves 1,905 tokens/s on average, demonstrating its effectiveness for long-context prefilling.
}

\begin{table}[t]
\centering
\caption{ Decode and prefill throughput (tokens/s) for \textsc{RapidMoE} against baselines tested with 1K sequence length. \textit{Long-Output} denotes decoding under reasoning workloads with an averaged output length at 8557.}
\label{tab:throughput}
\footnotesize  
\begin{tabular}{lccccccc}
\toprule
\multicolumn{1}{c}{\multirow{2}{*}{Stage}} & \multicolumn{1}{c}{\multirow{2}{*}{System}} & \multicolumn{6}{c}{Configuration, BatchSize} \\
\cmidrule{3-8}
 &  & C11,1 & C11,4 & C12,1 & C12,4 & C13,1 & C13,4 \\
\midrule
\multirow{4}{*}{Decode}
& llama.cpp & 5.4 & 5.7 & 5.3 & 5.8 & 9.7 & 10.0 \\
\cmidrule{2-8}
& KTrans. & 8.0 & 10.5 & 7.5 & 10.2 & 10.3 & 13.6 \\
\cmidrule{2-8}
\rowcolor{gray!10}
& \textbf{\textsc{RapidMoE}}
& \textbf{21.8} & \textbf{43.6}
& \textbf{20.5} & \textbf{42.7}
& \textbf{18.7} & \textbf{36.3} \\
 \rowcolor{gray!10}
& {\textit{Long-Output}}
& { \textbf{18.9}}
& { \textbf{37.8}}
& {--}
& {--}
& {\textbf{17.9}}
& {\textbf{34.8}} \\
\midrule
\multirow{3}{*}{Prefill} & llama.cpp & 17.7 & 18.2 & 17.9 & 18.7 & 60.0 & 61.4\\
\cmidrule{2-8}
& KTrans. & 62.8 & 62.8 & 61.3 & 61.3 & 102.5 & 102.5 \\
\cmidrule{2-8}
\rowcolor{gray!10}  &\textbf{\textsc{RapidMoE}} & \textbf{150.2} & \textbf{150.2} & \textbf{138.9} & \textbf{138.9} & \textbf{182.8} & \textbf{182.8} \\
\bottomrule
\end{tabular}
\end{table}

{ 
\subsubsection{Serving Performance}
\autoref{tab:serving} reports request-level TTFT, TPOT, and SLO attainment for serving workload specified in Sec.~6.1. 
At 0.2 requests/s with a concurrency of 4, \textsc{RapidMoE} meets the 100ms TPOT and 5s TTFT SLOs for most requests due to its substantially higher prefill and decode throughput compared to baselines.
These results demonstrate \textsc{RapidMoE}'s multi-request serving capability for low-concurrency local deployments.
}
\begin{table}[t]
\footnotesize
\centering
\caption{Serving performance over 1000 ShareGPT requests
at concurrency 4 and a request rate of 0.2\,rps. SLO attainment is
evaluated with $S_{\mathrm{TTFT}}=5$\,s and
$S_{\mathrm{TPOT}}=100$\,ms; Joint denotes the fraction of requests
satisfying both SLOs.}
\label{tab:serving}
\setlength{\tabcolsep}{4pt}
\begin{tabular}{
>{}c
>{}c
>{}c
>{}c
>{}c
>{}c}
\toprule
\multirow{2}{*}{Cfg.} &
\multirow{2}{*}{TTFT p50/p99} &
\multirow{2}{*}{TPOT p50/p99} &
\multicolumn{3}{c}{SLO Attainment (\%)} \\
\cmidrule{4-6}
& & & TTFT & TPOT & Joint \\
\midrule
\textsc{RapidMoE}-C11 & 3.41s/6.21s & 94ms/101ms & 93.9 & 92.3 & 87.4 \\
\textsc{RapidMoE}-C12 & 3.69s/6.37s & 97ms/112ms & 85.7 & 88.9 & 76.7 \\
\bottomrule
\end{tabular}
\end{table}
{
\subsubsection{Memory Efficiency}
\autoref{tab:matched_memory} details the memory breakdown for DeepSeek-V3 deployment at P1. 
With a comparable 149 GB peak HBM footprint, \textsc{RapidMoE} reduces peak DRAM usage to 240 GB, compared with 385 GB for KTrans. and 471 GB for MoE-APEX.
The storage reduction stems from the compact RESplit representation, which stores only Q2 Gate/Up residuals and Q4 Down weights on CPUs.
In contrast, KTrans. retains all Q4 expert weights, while MoE-APEX stores both low-precision (IQ1) and high-precision (Q4) copies to support adaptive precision.
Thus, the RESplit design reduces DRAM usage while delivering higher throughput as shown in Secs. \ref{sec6.3.1} and \ref{sec6.3.2}.
}

\subsection{Ablation Study}\label{sec6.4}

\subsubsection{Performance Breakdown}
\begin{figure}[t]
\centering
\includegraphics[width=0.5\textwidth,height=0.26\textwidth]{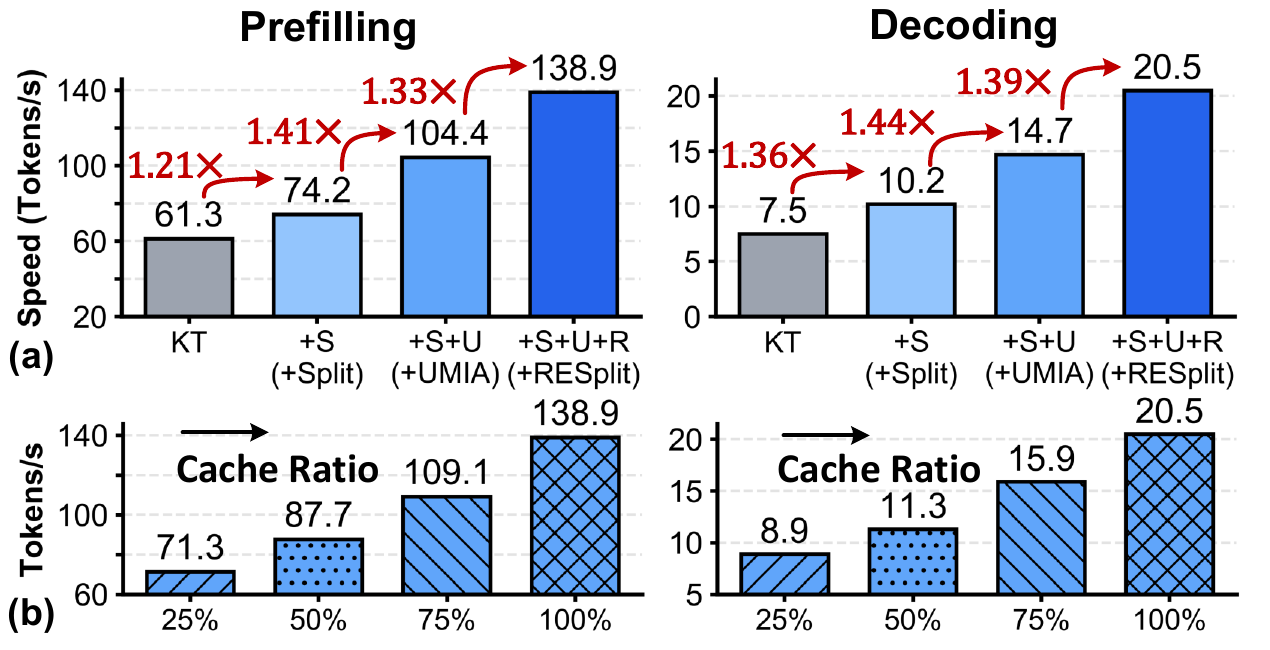}
  \caption{(a) Performance breakdown for prefilling and decoding phases on DeepSeek-V3. (b) Evaluation of different VRAM constraints, demonstrating inference speed when caching 25\% to 100\% of $W^Q$ on GPUs.}
  \label{fig:perf_mem}
\end{figure}

We ablate the system to quantify the effect of each component in \autoref{fig:perf_mem} (a). We start from baseline \texttt{KTransformers} (KT) and incrementally integrate our proposed modules as follows:

\noindent (1) \textbf{+ S}: Incorporating split routing with a static ($r=3$) yields a $1.2\times$--$1.3\times$ speedup. Statically partitioning the workload alleviates the fundamental CPU bottleneck.

\noindent (2) \textbf{+ S + U}: Dynamic arbitration provides an additional $\sim \! 1.4\times$ speedup. By navigating multi-level importance, UMIA minimizes the average CPU workload ($r$) without degrading model accuracy (shown in \autoref{fig:vis_r}).

\noindent (3) \textbf{+ S + U + R}: Deploying the RESplit representation and parallelism further transfers partial workload to GPUs and 
delivers a final speedup of up to $1.39\times$. 
\begin{table}[t]
\footnotesize
\centering
\caption{Memory breakdown at C12 configuration. HBM Other includes non-expert weights and CUDA graph/runtime buffers.
Peak denotes measured peak usage.}
\label{tab:matched_memory}
\setlength{\tabcolsep}{3pt}
\resizebox{\columnwidth}{!}{%
\begin{tabular}{
>{}l
>{}c
>{}c
>{}c
>{}c
>{}c}
\toprule
\multirow{2}{*}{\textbf{System}} &
\multicolumn{3}{c}{\textbf{HBM Breakdown (GB)}} &
\multicolumn{2}{c}{\textbf{DRAM Breakdown (GB)}} \\
\cmidrule(lr){2-4}\cmidrule(lr){5-6}
& Gate/Up/Down & Other & Peak
& Gate/Up/Down & Peak \\
\midrule

MoE-APEX
& 44/44/44 & 10+6 & 149
& 157/157/157 & 471 \\

KTrans.
& 42/42/51 & 10+5 & 150
& 120/120/145 & 385 \\

\rowcolor{gray!10}
\textbf{\textsc{RapidMoE}}
& 44/44/44 & 10+6 & \textbf{149}
& 63/63/113 & \textbf{240} \\

\bottomrule
\end{tabular}%
}
\end{table}

\subsubsection{Efficacy of UMIA}
\begin{figure}[t]
\centering
\includegraphics[width=0.5\textwidth,height=0.2\textwidth]{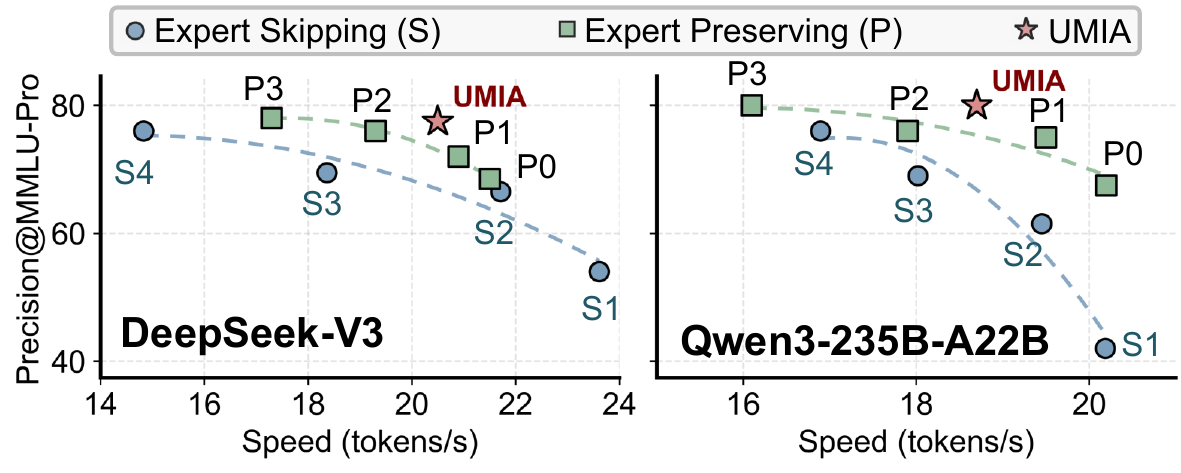}
  \caption{Accuracy-latency trade-off on (\textit{Left}) DeepSeek-V3 and (\textit{Right}) Qwen3-235B-A22B. Speed is evaluated using decoding throughput. S$x$ denotes statically skipping $(k-x)$ non-critical experts, while P$x$ denotes preserving them in low-precision, both using local router logits for arbitration.}
  \label{fig:umia}
\end{figure}

To validate UMIA, we compare it against two static routing baselines with a fixed critical set size $r$: (1) \textit{Skipping}, which discards the \(k-r\) non-critical experts, and (2) \textit{Preserving}, which retains them in low precision. As shown in \autoref{fig:umia}, \textit{Skipping} suffers from accuracy collapse. While \textit{Preserving} mitigates this loss, its fixed $r$ fails to adapt to importance divergence. 
UMIA adjusts $r$ based on multi-level runtime context, pushing the frontier forward with higher accuracy and throughput.

\begin{figure}[t]
\centering
\includegraphics[width=0.5\textwidth,height=0.25\textwidth]{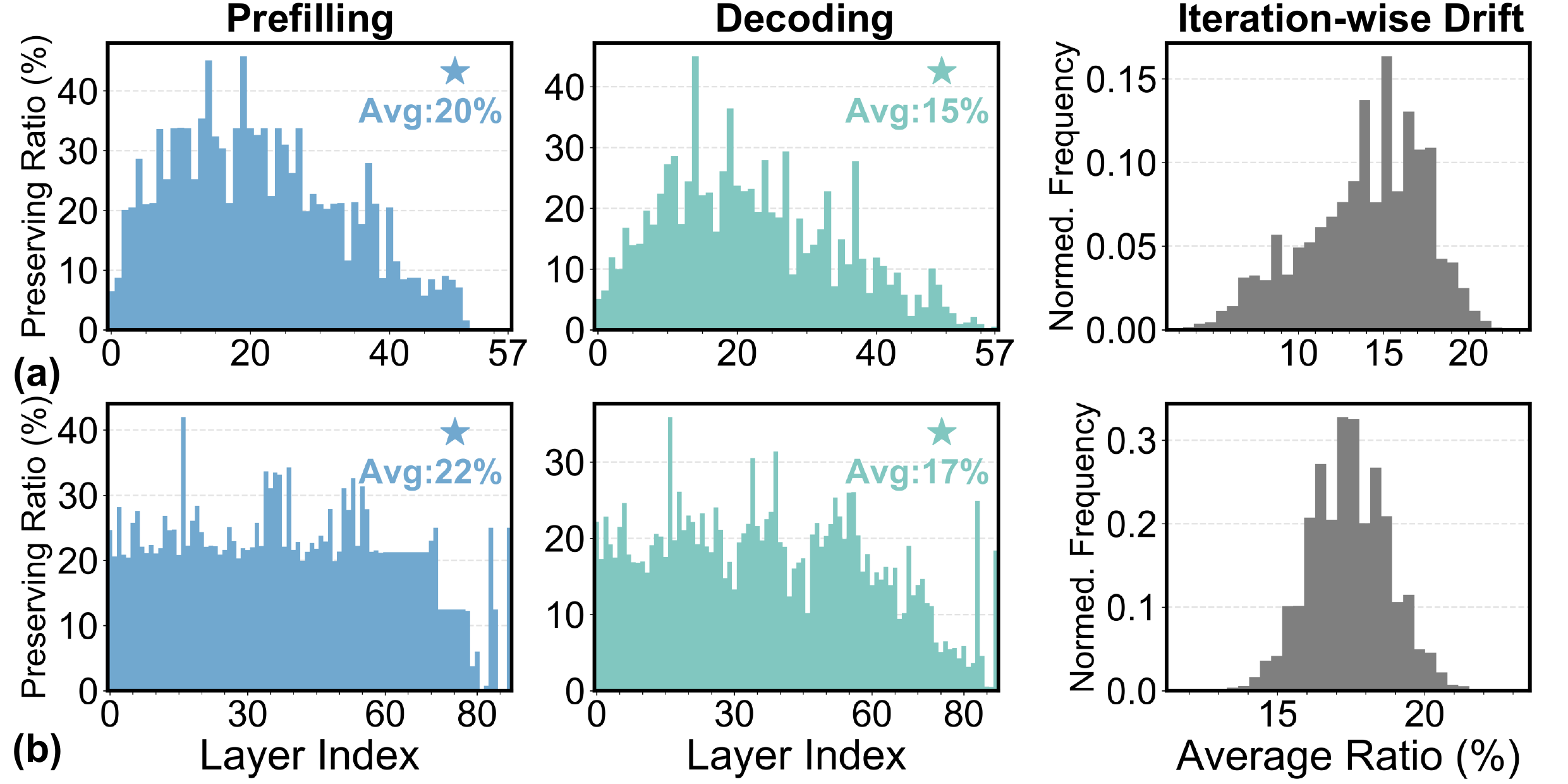}
  \caption{Visualization of preserving ratio ($r/k$) evaluated on MMLU-Pro for (a) DeepSeek-V3 and (b) Qwen3-235B-A22B. \textit{(Left \& Middle)} Averaged ratio across layers for prefill and decode phases. \textit{(Right)} Distribution across iteration steps.}
  \label{fig:vis_r}
\end{figure}

\noindent \textbf{Visualization Analysis.}
To demystify how UMIA achieves optimality, we visualize the distribution of critical set size in \autoref{fig:vis_r}, validating the multi-level divergence in Sec. \ref{sec3.1}.

\noindent (1-2) \textbf{Spatial and Temporal Divergence:} UMIA adaptively assigns a larger r to shallower layers while keeping $r$ minimal in deeper layers (spatial divergence). Temporally, prefilling exhibits a higher average $r$ than decoding, confirming its structural preservation of foundational KV-cache quality.

\noindent (3) \textbf{Iteration-wise Drift:} \autoref{fig:vis_r} also traces the real-time $r$ drift for different iteration steps. The observed dynamic shifts demonstrate that a fixed static $r$ cannot adapt to the inherent semantic drift of generated tokens over the inference process.

{
\begin{table}[t]
\footnotesize
\centering
\caption{Ablation of UMIA on DeepSeek-R1. All variants are calibrated
on MMLU-Pro; P/D denotes prefill/decode throughput (tokens/s).}
\label{tab:umia_ablation}
\setlength{\tabcolsep}{2.8pt}
\renewcommand{\arraystretch}{1.05}
\begin{tabular}{
>{}l
>{}c
>{}c
>{}c
>{}c
>{}c
>{}c}
\toprule
\textbf{Metric} &
\makecell{\textbf{static}\\$\boldsymbol{r}=3$} &
\makecell{\textbf{S-only}\\$\boldsymbol{r_{l}}$} &
\makecell{\textbf{T-only}\\$\boldsymbol{r_{t}}$} &
\makecell{\textbf{S+T}\\$\boldsymbol{r_{l,t}}$} &
\makecell{\textbf{L-only}\\ $\mathbf{g} > \Theta$} &
\cellcolor{gray!10}\textbf{UMIA} \\
\midrule
Acc.@AIME
& 73.3 & 60.0 & 66.7 & 63.3 & 66.7
& \cellcolor{gray!10}\textbf{73.3} \\

P/D@AIME
& 99/16.0 & 157/18.2 & 119/17.3 & 150/19.9 & 129/16.9
& \cellcolor{gray!10}\textbf{141/18.9} \\

P/D@MMLU
& 106/17.5 & 153/19.7 & 128/18.7 & 146/21.0 & 136/18.6
& \cellcolor{gray!10}\textbf{152/21.1} \\
\bottomrule
\end{tabular}
\end{table}
\noindent \textbf{Modulator Ablation.}
We ablate UMIA’s local (L), spatial (S), and temporal (T) modulators on DeepSeek-R1, calibrating their parameters on MMLU-Pro and evaluating accuracy on AIME-24. Static variants fix $r$ globally or along spatial (S-only) and temporal dimensions (T-only), while L-only retains only the input-dependent rule $\mathbf{g}>\Theta$. 
As shown in \autoref{tab:umia_ablation}, single-modulator variants degrade accuracy, while the static $r=3$ preserves accuracy at the cost of lower throughput. Full UMIA recovers the accuracy while sustaining competitive throughput, demonstrating the benefit of joint adaptation.
}
\subsubsection{Scalability to Varying VRAM Constraints}
We simulate varying VRAM constraints by capping the GPU cache capacity for low-precision experts ($W^Q$) in \autoref{fig:perf_mem} (b). We analyze the speed when caching 25\%-100\% of $W^Q$ with footprints of 37-149GB. 
Results show that \textsc{RapidMoE} sustains competitive throughput even under 25\% constraint, as it achieves speedups for hit experts and falls back to CPU-centric execution to handle missed experts.

Notably, while it scales to varying constraints, its performance sweet spot remains the regime where VRAM can hold the entirety of $W^Q$ to fully unlock the benefits. It is practical for real-world deployment, since low-bit quantization shrinks the $W^Q$ footprint of a 671B model to merely $\sim \! 150$GB, well within the capacity of typical single-node workstations.

\subsubsection{Latency Breakdown}
\begin{figure}[t]
\centering
\includegraphics[width=0.5\textwidth,height=0.29\textwidth]{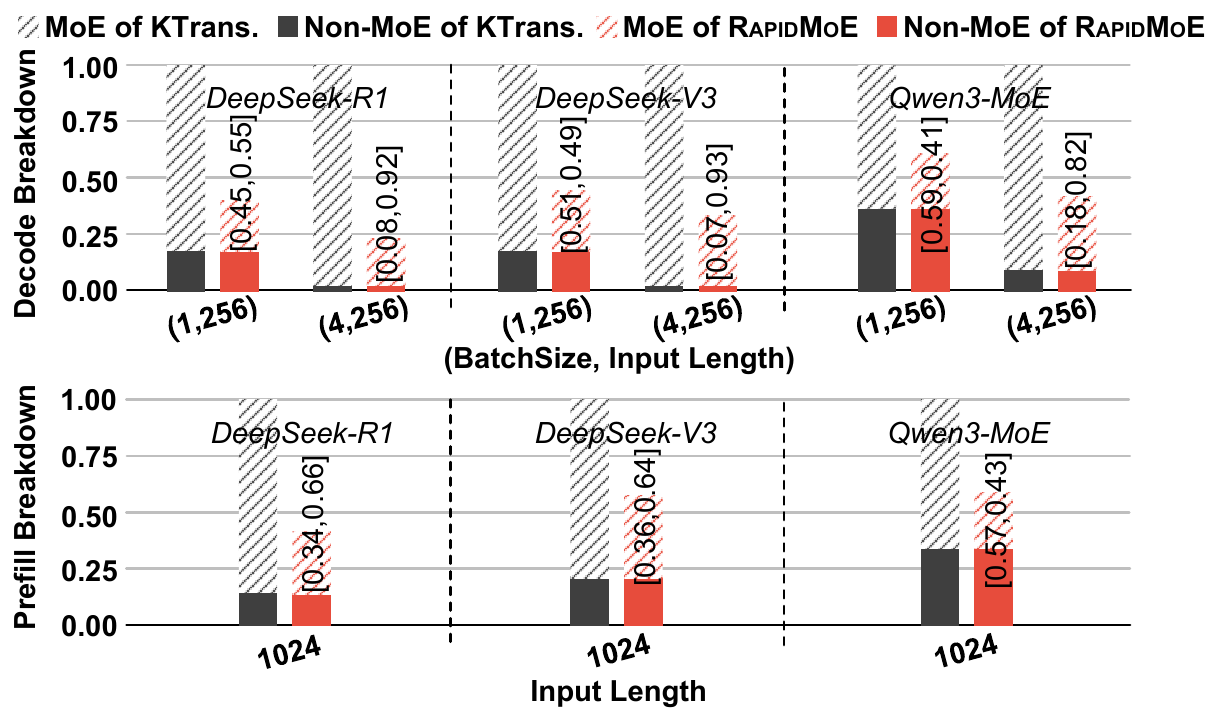}
  \caption{End-to-end latency breakdown of decode and prefill stage in KTransformers and \textsc{RapidMoE} at P1 platform.}
  \label{fig:breakdown}
\end{figure}

\autoref{fig:breakdown} presents the end-to-end latency breakdown.
The results reveal two insights. 
First, MoE operations consistently constitute the latency bottleneck in both stages.
Therefore, the optimization of MoE in \textsc{RapidMoE} significantly reduces end-to-end latency.
Second, implementing batch inference during decoding enhances the efficiency of operation executed on GPUs, thereby decreasing non-MoE proportion in overall latency.

{
\section{Discussion}
\label{sec:discussion}

\noindent \textbf{Memory--Accuracy--Latency Trade-off.}
\textsc{RapidMoE} exposes a deployment trade-off among memory, model accuracy, and inference latency. 
First, RESplit prioritizes full GPU residency of $W^Q$, whose format determines the memory--accuracy trade-off: lower precision reduces VRAM usage but increases quantization error. CPU residual computation then recovers errors without additional VRAM consumption.
Second, given a fixed $W^Q$ format, UMIA further balances accuracy and latency by dynamically adapting the critical set at runtime. Together, RESplit and UMIA address complementary dimensions of \textsc{RapidMoE}'s deployment trade-off.

\noindent \textbf{Deployment Scope.}
While fully GPU-resident inference is preferable when VRAM is sufficient, \textsc{RapidMoE} targets memory-constrained deployment, retaining the quantized path on GPUs while using CPU resources for selective residual correction. 
It enables efficient large-model inference with limited VRAM.
However, this hybrid execution also defines its concurrency scope: 
larger batches improve GPU utilization, but provide less benefit to CPU-bound residual computation.
\textsc{RapidMoE} is therefore best suited to low-concurrency, memory-constrained deployment, as higher concurrency increases TTFT and TPOT despite improving throughput.

\noindent \textbf{Hardware Generality.}
\textsc{RapidMoE} is validated on PCIe-connected Intel CPU--NVIDIA GPU systems. 
However, it does not fundamentally depend on specific hardware.
RESplit is built on hardware asymmetry arising from differences in compute capabilities and memory, which is expected to hold across most CPU-GPU platforms.
Implementation portability depends on kernel support.
Our GGUF-based GPU kernels build on llama.cpp~\cite{llama_cpp}, which provides CUDA, HIP, and Metal backends, while CPU kernels use common x86 extensions (AVX2) rather than Intel-specific features.
Supporting other platforms requires backend-specific adaptation, with the $W^Q$ format selected to fit the available VRAM. 

The primary hardware-dependent component is the asymmetric latency model in Sec.~\ref{4.2}.
Higher DDR bandwidth or CPU compute throughput, including advanced AMX with compatible kernels, can accelerate CPU-bound operations.
Likewise, higher GPU throughput or bandwidth benefits workloads limited by GPUs.
These improvements may alter the balance between CPU and GPU. 
Faster CPU execution shift the transition of $R(r)$ to larger $r$ by supporting more residual correction, whereas faster GPU execution reduces the overlap window for CPU and shifts it to smaller $r$.
Therefore, \(R(r)\) should be re-calibrated for each platform following \autoref{fig:timeline}, and achievable gains remain hardware-dependent.
}

\section{Related Work}
\noindent \textbf{Offloading Systems for MoE LLMs.} 
The memory and latency demands of MoE LLMs have spurred the development of offloading systems for memory-constrained devices \cite{moe-survey,deepspeed}. GPU-centric expert-offloading systems \cite{edgemoe,moe-infinity} execute MoE on GPUs with expert prefetching and caching to mitigate I/O overhead.
Recent advances \cite{moe-apex} optimize prefetching via layer-wise overlap and enhanced cache replacement policies, while \cite{moe-lightning} introduces novel pipelining for better throughput. 
However, these systems remain fundamentally limited by expert loading latency on the critical path, restricting scalability for large MoE models. Conversely, CPU-centric systems \cite{ktransformers,llama_cpp} leverage CPU computation for expert execution through optimized kernels when experts are unavailable on GPUs, addressing latency concerns.
While effective on advanced CPUs, such solutions under-utilize GPU resources, resulting in CPU bottlenecks and suboptimal throughput. 

\noindent \textbf{Model Compression for MoE LLMs.}
Model compression has emerged as a crucial direction for efficient LLM deployment \cite{model_compression}.
The key idea is to reduce memory footprint through methods such as pruning \cite{sparsegpt} and quantization \cite{quant_ori}.
For MoE LLMs, expert pruning \cite{expert-pruning1,expert-pruning2} reduces model size by systematically eliminating less important experts on calibration datasets. Quantization \cite{quant_ori} reduces parameter precision to decrease memory footprint \cite{quant1,smoothquant}. Widely adopted quantization methods including AWQ \cite{awq}, GPTQ \cite{gptq}, and GGUF \cite{gguf} have been integrated into mainstream inference engines (e.g., vLLM \cite{vllm}, SGLang \cite{sglang}, llama.cpp \cite{llama_cpp}). Our work builds upon GGUF due to its superior cross-platform compatibility. In contrast, alternatives like AWQ and GPTQ offer limited CPU support. This choice enables consistent performance across heterogeneous devices.

\section{Conclusion}
We propose \textsc{RapidMoE}, an adaptive expert-residual offloading system for efficient large-scale MoE inference on heterogeneous platforms. By exploiting cross-asymmetry in algorithmic routing and hardware characteristics, \textsc{RapidMoE} employ systematic residual-split framework with a novel runtime arbitration to achieve optimal resource utilization, latency and throughput. Experiments demonstrate its practicality in real-world deployments, achieving 43 tokens/s in decode and 150 tokens/s in prefill for DeepSeek-R1-671B on 2×A800-80G GPUs. Overall, it represents a significant advancement in efficient deployment of large-scale MoE in memory-constrained environments.

\begin{acks}
We sincerely thank anonymous reviewers and our shepherd for their feedback and constructive suggestions.
This work was supported by the Fundamental and Interdisciplinary Disciplines Breakthrough Plan of the Ministry of Education of China (JYB2025XDXM120), Beijing National Research Center for Information Science and Technology (BNRist).
\end{acks}

\newpage
\bibliographystyle{ACM-Reference-Format}
\bibliography{acmart}

\end{document}